\documentclass[twocolumn,floats,floatfix,showpacs,amssymb,prd,superscriptaddress,nofootinbib]{revtex4-1}
\usepackage{graphicx, epsfig, amssymb} 
\usepackage{amsmath, amsfonts}
\usepackage{bm}

\usepackage[utf8]{inputenc}

\usepackage[linktocpage]{hyperref}
\usepackage[caption=false]{subfig}
\usepackage[usenames]{xcolor}

\usepackage{enumerate}

\usepackage[normalem]{ulem}

\def\Lie{\mathcal{L}}
\def\p{\partial}

\renewcommand{\vec}[1]{\boldsymbol{#1}}

\def\be{\begin{equation}}
\def\ee{\end{equation}}
\newcommand{\beq}{\begin{eqnarray}}
\newcommand{\eeq}{\end{eqnarray}} 
\newcommand{\ba}{\begin{align}}
\newcommand{\ea}{\end{align}}

\begin{document}

\title{\large Black hole binaries:\\ ergoregions, photon surfaces, wave scattering, and quasinormal modes}

\author{Thiago Assump\c c\~ao}
\affiliation{Centro de Matem\'atica, Computa\c c\~ao e Cogni\c c\~ao,
Universidade Federal do ABC (UFABC), 09210-170 Santo Andr\'e, S\~ao Paulo, Brazil}

\author{Vitor Cardoso}
\affiliation{CENTRA, Departamento de F\'{\i}sica, Instituto Superior
T\'ecnico, Universidade de Lisboa,
Avenida Rovisco Pais 1, 1049 Lisboa, Portugal}
\affiliation{Perimeter Institute for Theoretical Physics, Waterloo, Ontario N2L 2Y5, Canada}
\affiliation{Theoretical Physics Department, CERN, CH-1211 Geneva 23, Switzerland}

\author{Akihiro Ishibashi}
\affiliation{Department of Physics and Research Institute for Science and Technology, Kindai University, Higashi-Osaka, 577-8502, Japan}

\author{Maur\'icio Richartz}
\affiliation{Centro de Matem\'atica, Computa\c c\~ao e Cogni\c c\~ao,
Universidade Federal do ABC (UFABC), 09210-170 Santo Andr\'e, S\~ao Paulo, Brazil}

\author{Miguel Zilh\~ao}
\affiliation{CENTRA, Departamento de F\'{\i}sica, Instituto Superior
T\'ecnico, Universidade de Lisboa,
Avenida Rovisco Pais 1, 1049 Lisboa, Portugal}

\begin{abstract}
Closed photon orbits around isolated black holes are related to important aspects of black hole physics,
such as strong lensing, absorption cross section of null particles and the way that black holes relax through quasinormal ringing.
When two black holes are present -- such as during the inspiral and merger events of interest for gravitational-wave detectors -- the concept of closed photon orbits still exist, but its properties are basically unknown. With these applications in mind, we study here the closed photon orbits of two different {\it static} black hole binaries.
The first one is the Majumdar-Papapetrou geometry describing two extremal, charged black holes in equilibrium, while the second one is the double sink solution of fluid dynamics, which describes (in a curved-spacetime language) two ``dumb'' holes. For the latter solution, we also characterize its dynamical response to external perturbations, and study how it relates to the photon orbits.
In addition, we compute the ergoregion of such spacetime and show that it does {\it not} coincide with the event horizon. 
\end{abstract}


\maketitle

\section{Introduction}
\label{sec:intro}
High-frequency electromagnetic or gravitational waves propagating around extremely compact objects -- such as black holes (BHs) -- can travel in closed orbits. These can be either stable or unstable, according to whether or not small deviations from the orbit grow away from it as time evolves. In the case of a Schwarzschild BH, the union of all such orbits forms a sphere, the so called Schwarzschild photon sphere. Any inward-directed lightray emitted inside the photon sphere will eventually end in the singularity, while any outward-directed lightray emitted outside eventually reaches future null infinity. Hence the photon sphere is not a horizon (for which {\it no} lightray would ever be able to exit) but has a similar fundamental role.

Several BH phenomena are directly related to the presence of a photon sphere~\cite{Ferrari:1984zz,Cardoso:2008bp,Yang:2012he,Berti:2009kk,Cornish:2003,Decanini:2011xi,1993AmJPh..61..619N,Virbhadra:1999nm,Keir:2014oka,Cardoso:2014sna,Cunha:2017qtt,Cardoso:2016rao,Cardoso:2017cqb}. First, the characteristic modes of oscillation of a spherically symmetric BH can be related to the parameters of the null closed orbits~\cite{Ferrari:1984zz,Cardoso:2008bp, Yang:2012he,Berti:2009kk,Glampedakis:2017dvb,Pappas:2018opz}. In the eikonal limit, the oscillation frequency of such modes is a multiple of the orbital frequency while their characteristic damping time is related to the instability timescale of such orbits~\cite{Cornish:2003,Cardoso:2008bp,Yang:2012he,Berti:2009kk}. Second, the absorption cross section of a highly energetic scalar field scattered by the gravitational field of a spacetime endowed with a photon sphere is simply described by the orbital frequency and instability timescale of the associated null closed orbits~\cite{Decanini:2011xi}. 
Third, the extreme bending of light near the photon sphere also allows the formation of multiple (possibly infinite) images on the optical axis~\cite{1993AmJPh..61..619N,Virbhadra:1999nm} - in addition to the primary image typically formed by any weak gravitational lens. In particular, strong gravitational lenses produce an infinite number of Einstein rings, and the astrophysical observation of a sequence of such rings would be another successful test of General Relativity in the strong field regime~\cite{Virbhadra:1999nm}. Last but not least, the presence of stable photon spheres implies the existence of a trapping region in spacetime where matter can accumulate, potentially leading to nonlinear instabilities~\cite{Keir:2014oka,Cardoso:2014sna,Cunha:2017qtt,Cunha:2017eoe}.

The concept of a photon sphere, useful for Schwarzschild BHs, can be generalized to the concept of a photon surface in an arbitrary spacetime~\cite{2001JMP....42..818C}. The definition of a photon surface, in simple terms, requires that any null geodesic tangent to the surface will remain everywhere tangent to it. More specifically, in this work our main interest is in closed null hypersurfaces, i.e.~non-spacelike, chronal hypersurfaces $S$ that admit a null geodesic (i) which is tangent to S and (ii) whose projection onto a given spatial cross-section of S forms a ``closed orbit" on the cross-section~\footnote{Note that the ``closed orbit" here defined is not the same as a closed geodesic}.
The notion of a photon surface is independent of any symmetry that the spacetime may possess, and is applicable to the situation of two very compact objects that inspiral into each other and eventually merge, forming a single compact object. These events, notably the merger phase, being important sources of gravitational-waves, are of interest for detectors like LIGO and LISA. 
Photon surfaces and closed photon orbits around such binaries, although never studied before, could provide important astrophysical information about BH-BH (and BH-neutron star) collisions, specially now that the gravitational-wave astronomy era has begun~\cite{Barack:2018yly}.  

The main objective of this work is to provide the first step towards understanding closed photon orbits around binary systems of astrophysical compact objects. Due to the complexity of simulating and studying collisions of BHs and other compact objects, we focus on simpler geometries of BH binaries that are static and, therefore, never merge. We expect that our results are useful for low-frequency binaries whose velocities are much smaller than the speed of light. This work is also a first step towards understanding how other peculiarities -- such as ergoregions -- extend 
to dynamical spacetimes. There is a clear motivation for that, since it could lead to Penrose-like phenomena or interesting superradiant effects~\cite{Brito:2015oca}.

The first toy model we consider is that of two electrically charged BHs in equilibrium~\cite{Majumdar:1947eu,Papapetrou:1947a,Hartle:1972ya} (the Majumdar-Papapetrou solution of Einstein's equations), for which we study null geodesics both on the symmetry plane between the BHs and on the plane containing both BHs. Our second model is based on analogue gravity, the fact that sound propagation in moving fluids is formally equivalent to a scalar field evolving in curved spacetime~\cite{1981PhRvL..46.1351U,Barcelo:2005fc}. Points where the fluid velocity exceeds the local sound speed behave as an ergoregion; if this happens on a normal to a certain surface, such surface is an analogue horizon.
By considering a double sink hydrodynamical fluid flow (an exact solution of the fluid-dynamic equations) we are able to study the analogue of a binary BH system. 
We should note that extending the notion of ergoregions to dynamical curved spacetimes is nontrivial in general, but quite straightforward in acoustic setups.
Our results may thus help in understanding whether phenomena such as superradiance exist in BH-binary spacetimes~\cite{Brito:2015oca}~\footnote{The existence of ergoregions in
static spacetimes describing two Kerr BHs held apart by a strut was established in Ref.~\cite{Costa:2009wj}. A generalization to dynamical regular spacetimes has not been done.}.

We note that partial results concerning the structure of lightray or scalar-field propagation in the background of BH binaries were reported earlier. The evolution of scalar fields in the background of an inspiralling BH binary was investigated in Ref.~\cite{Okounkova:2017yby}. The shadow of a static BH binary (unphysical, since it contains a strut holding the two BHs apart) was also recently reported~\cite{Cunha:2018gql,Cunha:2018cof}. However, unlike our analysis, these references do not address the question of the structure and timescales associated to the different possible closed photon surfaces. Therefore, through our work, we hope to pave the way for future investigations of closed null orbits in more realistic situations of interest for LIGO and LISA.

\section{The Majumdar-Papapetrou solution}
%
\begin{figure}[ht]
\includegraphics[width=0.45\textwidth]{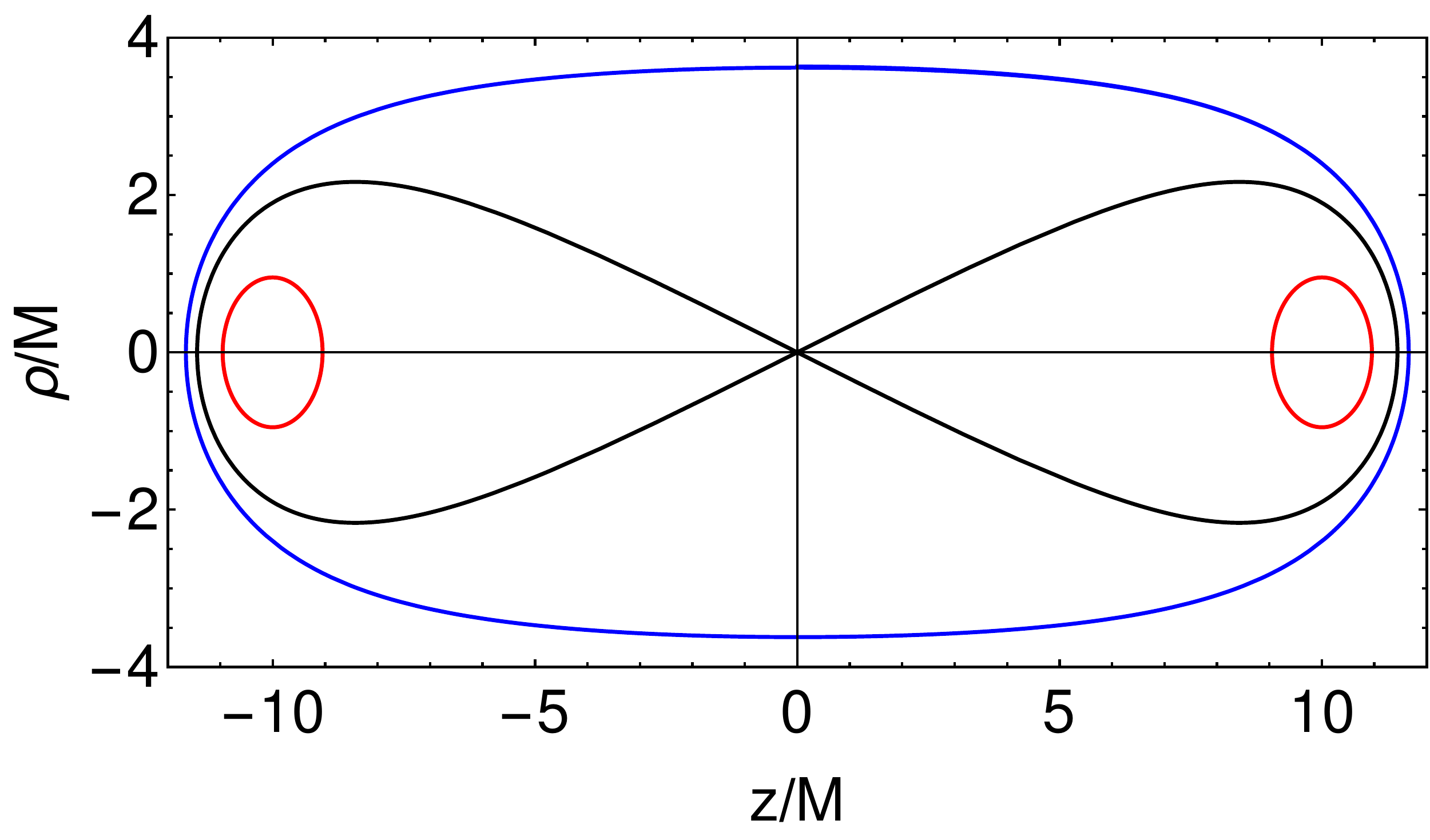}
\caption{There are different closed orbits for light in a static binary. We find, generically, three different closed trajectories: one global outer geodesic that encircles both holes, an ``8-shaped'' trajectory and two ``smaller'' ones encircling only each of the BHs.
These plots show the different ``light rings'' for a MP solution with $a=10M$. The ``8-shaped'' orbit (black curve) has a period
$T=98.79M$ in coordinate time.
The global outer orbit (blue curve) has a period $T=95.47M$. 
The period of the orbits around a single hole (red curves) is $T=26.39M$.
\label{fig:null_geo_phi0_MP}}
\end{figure}
\begin{figure}[ht]
\includegraphics[width=0.45\textwidth]{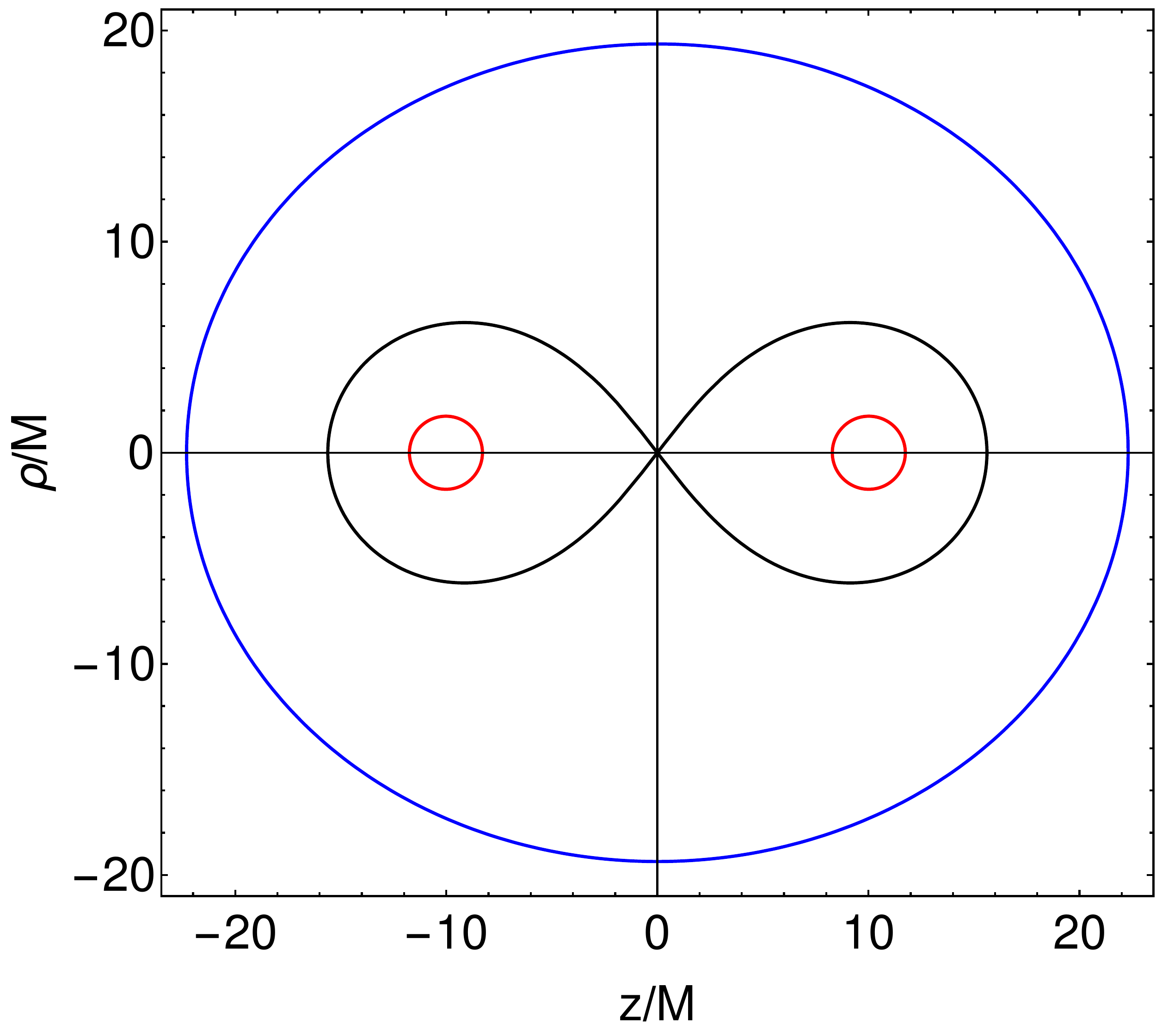}
\caption{Same as Fig.~\ref{fig:null_geo_phi0_MP}, but for massive free particles instead of light (i.e.~timelike instead of null geodesics). The ``8-shaped'' orbit  (black curve) has energy
$E=0.9159$ and period $T=291.64M$ in coordinate time.
The global outer orbit (blue curve), has energy $E=0.9644$ and period $T=474.01 M$.
The orbits around a single hole (red curves) have energy $E=0.9159$ and period $T=38.84 M$.
\label{fig:timelike_geo_phi0_MP}}
\end{figure}
An exact solution in General Relativity which describes two or more static BHs is known as the Majumdar-Papapetrou (MP) solution~\cite{Majumdar:1947eu,Papapetrou:1947a,Hartle:1972ya}. In a cylindrical coordinate system the two-BH version of the MP solution is written as
\be
\label{eqn:MPmetric}
ds^2=-\frac{dt^2}{U^2}+U^2\left(d\rho^2+\rho^2d\phi^2+dz^2\right)\,,
\ee
with 
\be
\label{eqn:MP_U}
U(\rho,\,z)=1+\frac{M}{\sqrt{\rho^2+(z-a)^2}}+\frac{M}{\sqrt{\rho^2+(z+a)^2}}\,.
\ee
This solution represents two maximally charged BHs in equilibrium, each with mass $M$ and charge $Q=M$. Here, and throughout this work, we use geometric units. In these coordinates, their horizons are shrunk to two points at $z= \pm a$ (hence, the parameter $a$ measures the distance between them). The spacetime ADM mass is $2M$.

The Lagrangian for geodesic motion in the Majumdar-Papapetrou spacetime is given by
\begin{equation}
\label{eqn:lagrangian_MP}
2 \mathcal{L} = - \frac{\dot{t}^2}{U^2} + U^2(\dot{\rho}^2 + \rho^2 \dot{\phi}^2 + \dot{z}^2) = - \delta \,,
\end{equation}
where dots refer to derivatives with respect to an affine parameter. The constant $\delta$ on the right handside is zero ($\delta=0$) for null geodesics and one ($\delta=1$) for timelike geodesics. The two conserved quantities are $E=\dot{t}/U^2$ and $L=\dot{\phi}U^2\rho^2$, which we identify as energy and angular momentum per unit rest mass, respectively.
 
\subsection{Geodesics in the meridian $\phi=0$ plane}
We can fix the azimuth angle by setting $\phi=0$ (this is an arbitrary choice as any fixed angle yields the same behavior) and study geodesic motion confined to the meridian plane. For a generic separation between the BHs, the geodesic equations are not separable, as noted in Ref.~\cite{Chandrasekhar:1989vk}.
However, when the two BHs coalesce, $a = 0$ and it is easy to solve the corresponding geodesic equations~\cite{Cardoso:2008bp}. We find that there is a null geodesic at $r=\sqrt{z^2+\rho^2}=2M$. 

With the aid of standard numerical integration methods, three types of closed null and closed timelike orbits were identified, as shown in Figs.~\ref{fig:null_geo_phi0_MP} and \ref{fig:timelike_geo_phi0_MP}:
\begin{enumerate}[{\bf i.}]
\item An orbit which encloses both horizons, shown in blue in both figures.  For
  {\it small} separations, i.e.~$a \ll M$, the coordinate radius of such an
  orbit is $2M$, as expected from the analysis of the $a = 0$ case. The
  corresponding coordinate period is $\sim 16 \pi M$. At {\it large}
  separations, we find that the period of null outer orbits is
  $T \sim 20 \pi M + 4a$, for values of $M$ in the range
  $1\lesssim M \lesssim 10$.

\item Orbits which enclose only one horizon, shown in red in both figures.  At
  {\it large} separations, such geodesics only ``see'' the gravitational field
  of one BH. Indeed, we find that then the orbit is nearly circular with radius
  $\sim M$, and a coordinate period $\sim 8 \pi M$, as expected from the above.

\item An ``8-shaped'' orbit which encloses each of the horizons once, shown in
  black in both figures.

\end{enumerate}

For the two-BH MP geometry these three types of orbits seem to exist for any separation $a$. These results are in complete agreement with both Chandrasekhar's and Shipley and Dolan's recent results~\cite{Chandrasekhar:1989vk,Shipley:2016omi}.    
A full description of the equations and normalization conditions is given in Appendix~\ref{app:geoMP}. We found, numerically, that all these null geodesics are unstable: the numerical integration requires fine tuning to follow them. On the other hand, timelike geodesics are more stable and complete a larger number of full orbits before collapsing to the singularity (some ``8-shaped'' trajectories are actually long-term stable). For highly relativistic particles, timelike geodesics approach null geodesics. 
It is impossible to find closed timelike orbits for arbitrary energies.

\subsection{Geodesics in the symmetry $z=0$ plane}
%
\begin{figure}[ht]
\begin{tabular}{ccc}
\includegraphics[width=0.15\textwidth,clip]{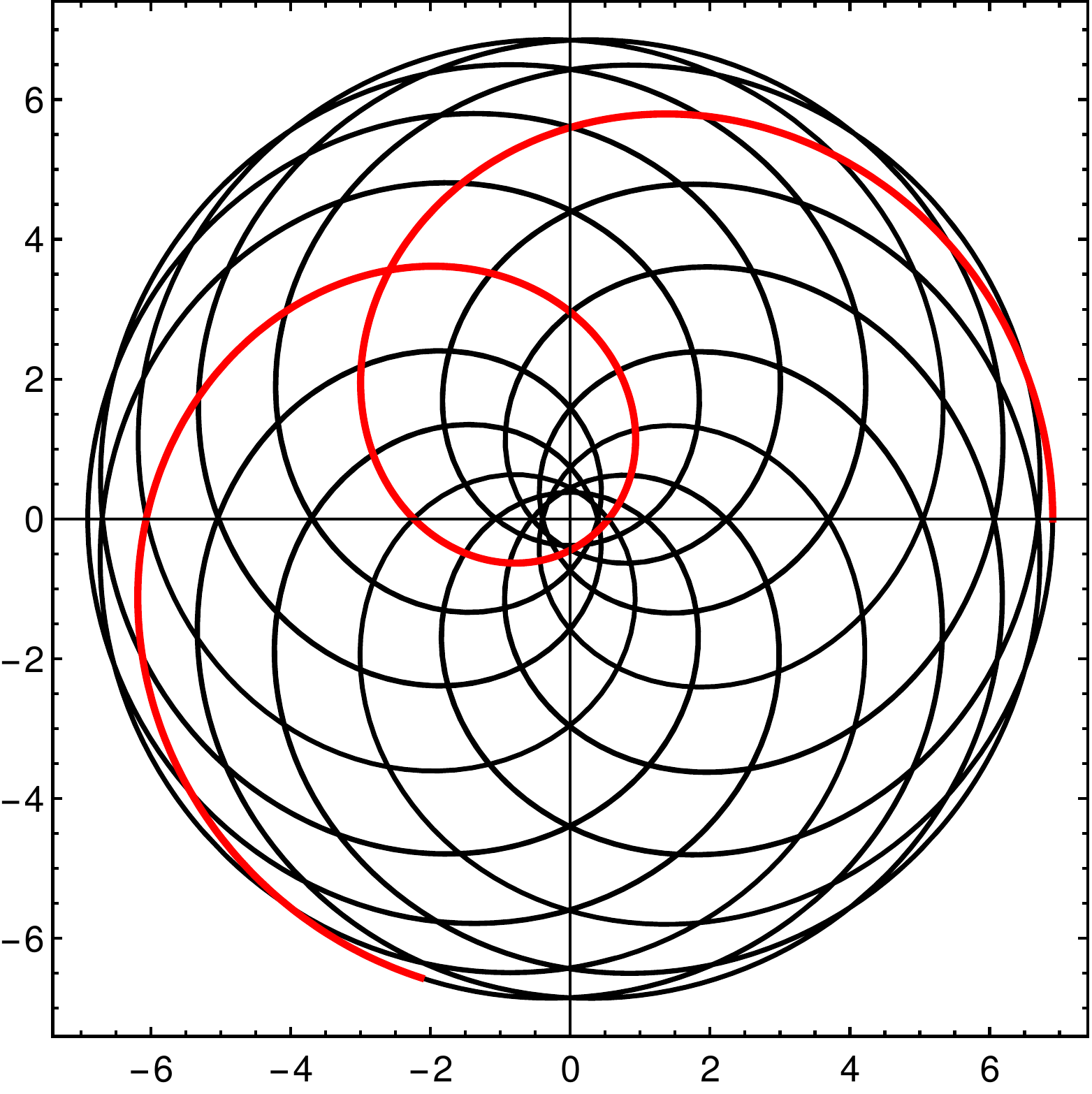}&
\includegraphics[width=0.15\textwidth,clip]{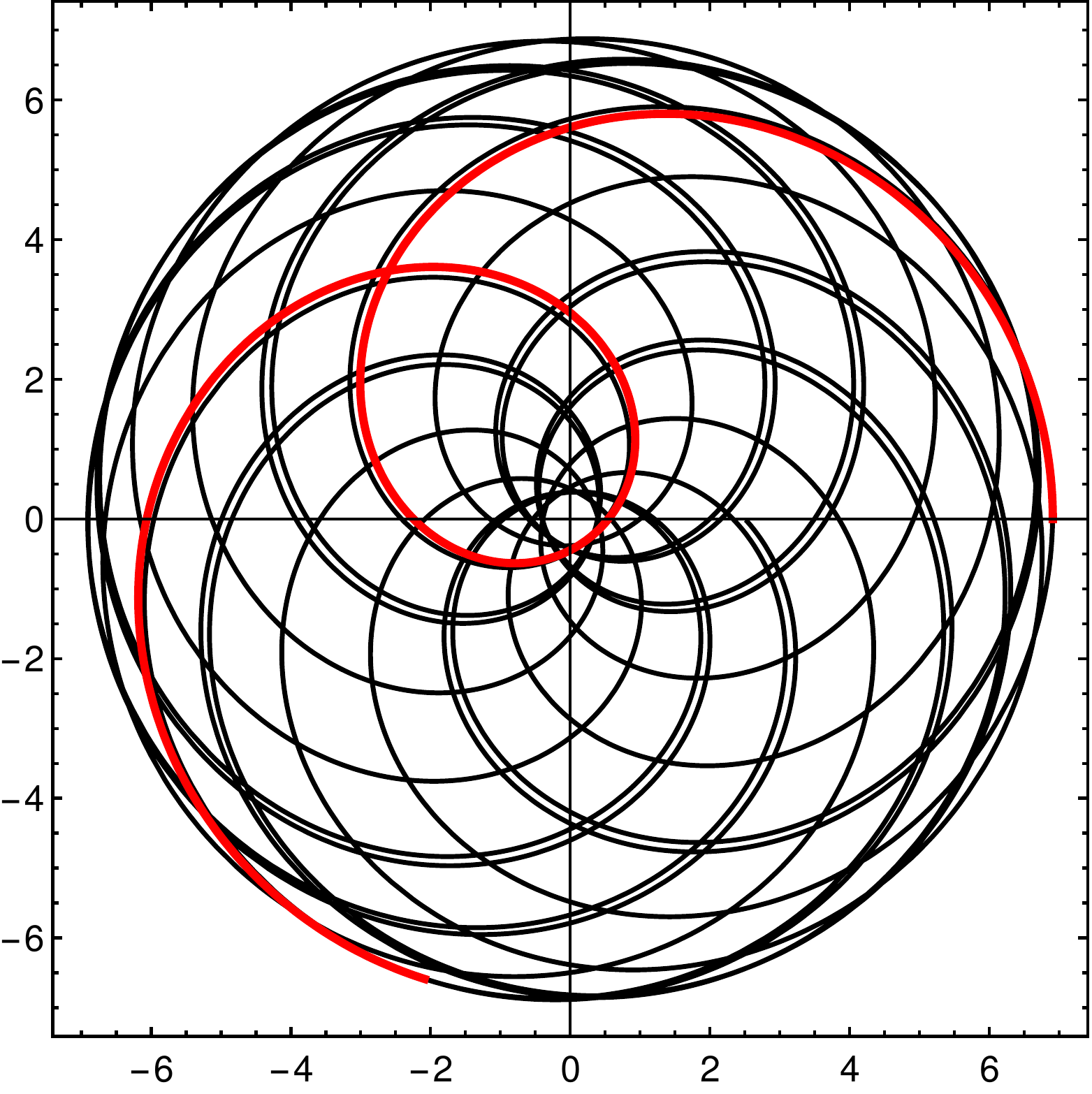}&
\includegraphics[width=0.15\textwidth,clip]{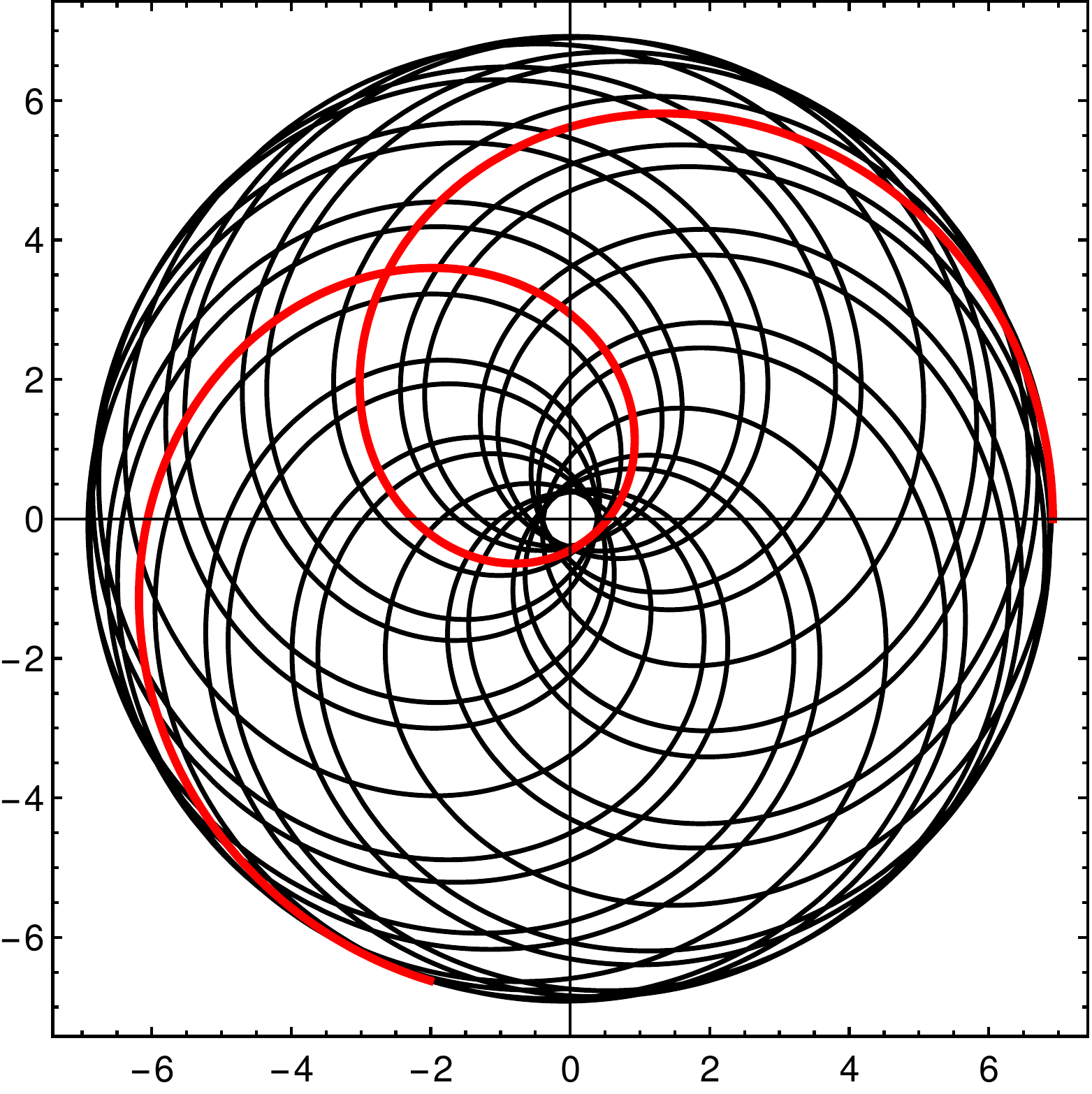}
\end{tabular}
\caption{Null geodesics on the symmetry plane of a two-BH MP spacetime, characterized by $a=M/5$.
The radii of the circular orbits do not depend on $L$, and are expressed in units of $a$. 
From left to right, the energy of the orbits are: $E=0.0366882, 0.0366939, 0.0367020$. The red curve shows the portion of the motion between two consecutive apastrons, for which the accumulated angle is given by $\Delta \phi = 2 \pi q$. When $q$ is a rational number, the motion is periodic, and the geometric properties of the plot can be extracted from $q$. From left to right, $q$ is given by $(1+7/10)$, $(1+33/47)$, and $(1+45/64)$. Notice that $q$ increases monotonically with $E$. Geometrically this means that the motion is divided into more ``leaves", which creates more vertices in the figures.  
The stable circular orbit has a radius $\rho_{\rm in}=1.167a$, and the unstable orbit is at $\rho_{\rm out}=9.740a$.
\label{fig:periodic1}
}
\end{figure}
\begin{figure}[ht]
\begin{tabular}{c}
\includegraphics[width=0.45\textwidth]{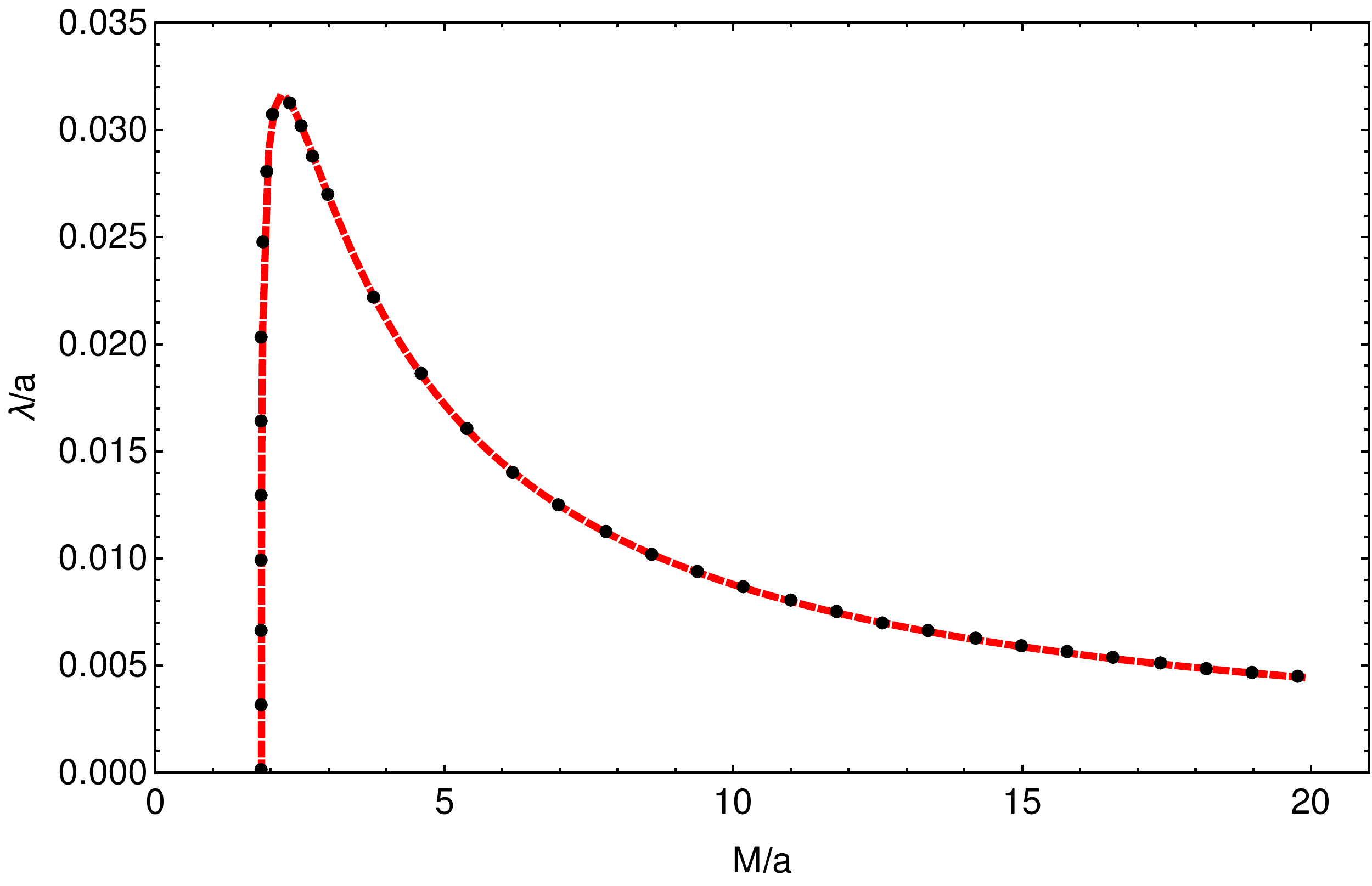}
\end{tabular}
\caption{Lyapunov exponent as a function of the BHs masses for null geodesics in the $z=0$ plane of the MP spacetime. The black dots are the numerical results, while the red dashed line corresponds to the analytic expression.\label{lyapunov_exponent1}}
\end{figure}
\begin{figure}[ht]
\begin{tabular}{ccc}
\includegraphics[width=0.158\textwidth,clip]{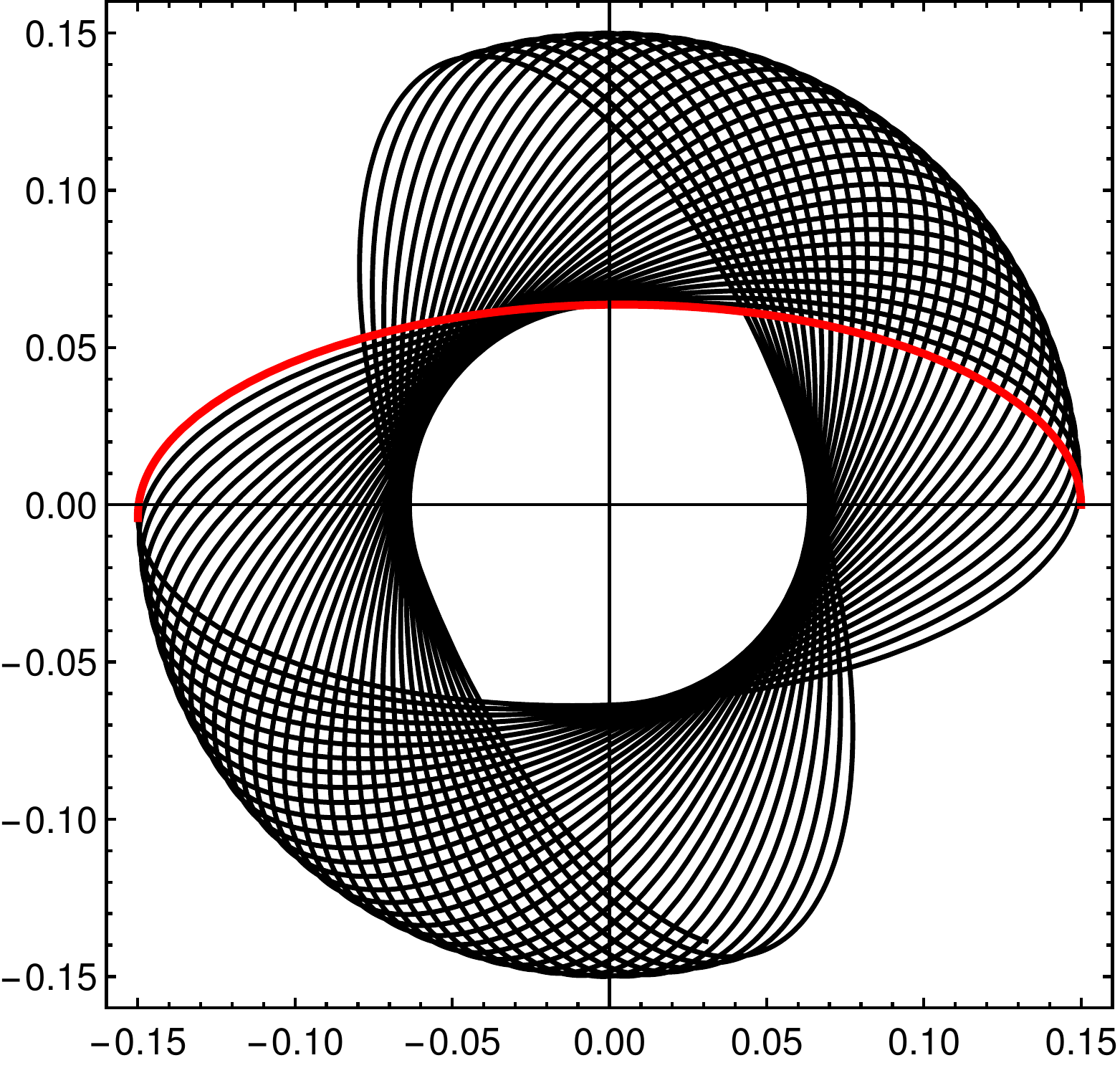}&
\includegraphics[width=0.15\textwidth,clip]{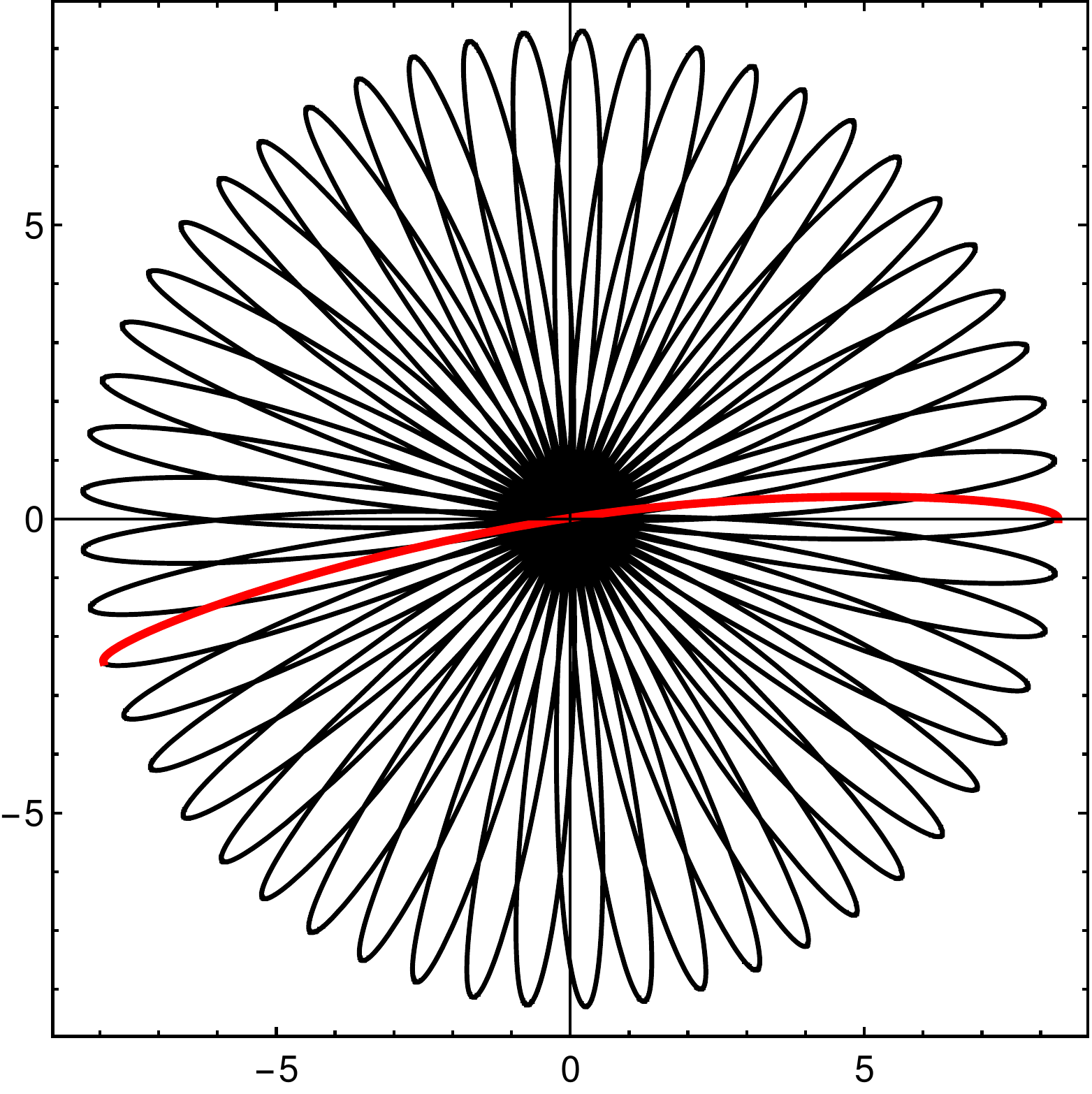}&
\includegraphics[width=0.154\textwidth,clip]{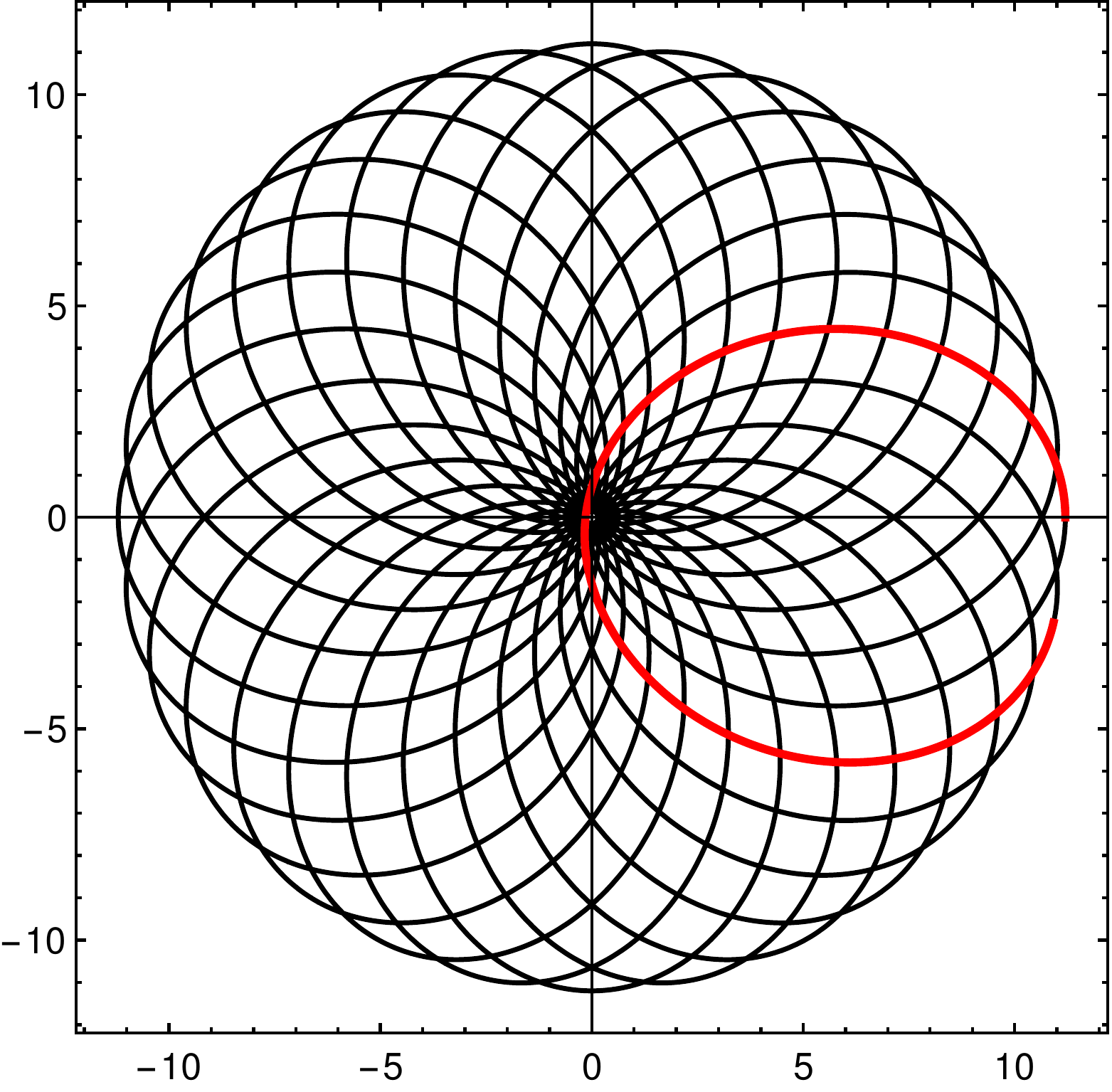}
\end{tabular}
\caption{Timelike geodesics on the symmetry plane of a two-BH MP spacetime, characterized by $a=M/5$. From left to right, the energy and angular momentum of the orbits are: 
$(E=0.0920047, L=0.1a)$, $(E=0.456023, L=a)$, and $(E=0.585419, L=10a)$. 
From left to right, $q$ is given by $56/111$, $29/53$, and $27/28$. The radii of both circular orbits (stable and unstable) depend on $M$ and $L/a$ for timelike geodesics. From left to right, the inner circular orbit radii $\rho_{\rm in}/a$ are given by $0.0977, 0.3093, 0.8892$. There are no unstable outer orbits for these combinations of $M$ and $L/a$.
\label{fig:periodic2}
}
\end{figure}
Half way between the two BHs, there is a symmetry plane ($z=0$) in which the geodesic equations are separable. By setting  
$a=1$, which simply amounts to rescaling units as $\rho \rightarrow \rho /a$ and $M \rightarrow M /a$, Eq.~\eqref{eqn:MP_U} becomes
\be
U = 1+ \frac{2M}{\sqrt{\rho^2 + 1}}\,.
\ee
Substituting $E$ and $L$ and setting $\delta =0$ in the normalization condition, the equation for null geodesics becomes
\beq
\dot{\rho}^2 + V_{\rm eff}(\rho) &=& E^2\,,\\
V_{\rm eff}(\rho) &=& \frac{L^2}{\rho^2 U^4}\,.\label{eqn:null_radial}
\eeq

Provided that $M^2>27/8$, the effective potential $V_{\rm eff}$ has two critical points, which correspond to a stable (inner) and an unstable (outer) closed orbit.\footnote{These statements refer to stability on the $z=0$ plane only. In reality, both orbits are expected to be globally unstable: any perturbation on the $z$ direction would give rise to motion towards one of the holes.} The analytic expressions for these two radii are~\cite{Patil:2017}
\beq
\rho_{\rm out} &=& \sqrt{\frac{4M^2}{9}\left(1+2 \cos\left[\frac{1}{3}\cos^{-1}\gamma \right] \right)^2-1} \,, \nonumber \\
\rho_{\rm in} &=& \sqrt{\frac{4M^2}{9}\left(1-2 \sin\left[\frac{\pi}{6}-\frac{1}{3}\cos^{-1} \gamma \right] \right)^2-1}\,,\nonumber
\eeq
where $\gamma\equiv 1- 27/(4M^2) $.
Notice that $\rho_{\rm in}^2 = 1 + \sqrt{2}/ M + {\cal O}(1/M^2)$, and $\rho_{\rm out}^2 = -5 + 4M^2 +{\cal O}(1/M^2)$. When $M \gg a$ we have $\rho_{\rm out} \rightarrow 2M$, so that the single hole solution is recovered. Besides these circular orbits, more general closed photon orbits also exist in the symmetry plane, as shown in Fig.~\ref{fig:periodic1}. There is, however, no apparent connection between these closed orbits and the ones found in the meridian plane.

For timelike geodesics, the number of roots of the equation $d V_{\rm eff} / d \rho = 0$ (and consequently the number of circular orbits) depends not only on the mass of the BHs but also on the angular momentum of the particle due to an extra term in the effective potential~\cite{circular_orbits_RN}. The Lyapunov exponent $\lambda$ associated with the unstable circular orbits can be computed in a straightforward way~\cite{Cornish:2003,Cardoso:2008bp}. A detailed calculation for null geodesics is given in Appendix~\ref{app:geoMP}. In our context, the Lyapunov exponent $\lambda$ describes the time scale in which a deviation $\delta \rho$ from a circular orbit grows exponentially, i.e.~$\lambda$ is characterized by $\delta \rho \sim e^{\lambda t}$.
As seen in Fig.~\ref{lyapunov_exponent1}, the Lyapunov exponent attains its maximum value $0.03158$ at $M=2.197$, and approaches $0$ for very large $M$. For instance, at $M=10^6$, the exponent is $8.38 \times 10^{-8}$. More general closed timelike orbits -- other than circular -- also exist, as shown in Fig.~\ref{fig:periodic2}.

Closed orbits in the symmetry plane exhibit a ``zoom-whirl'' behavior, as seen in Figs.~\ref{fig:periodic1} and~\ref{fig:periodic2}. A brief remark concerning these orbits is in order. Applying the formalism presented in Ref.~\cite{periodic_orbits2008}, we associate geometric properties of symmetric orbits in the plane $z=0$ with the accumulated angle between two consecutive apastrons $\Delta \phi$ given by
\be
\Delta \phi = 2 \int_{\rho_p}^{\rho_a} \frac{d \phi}{d\rho} d\rho \,,
\ee
where $\rho_p$ and $\rho_a$ are the radial coordinates of the periastron and apastron respectively.
We may look for orbits for which $\Delta \phi = 2 \pi q$, where $q$ is a rational number. We can always write any non-integer rational number uniquely as $q=w + v/z$, where $w=\left \lfloor q \right \rfloor \in \mathbb{Z}$, and $v<z$ are coprime positive integers.
 The whirl number $w$ gives the number of times the orbit whirls between two consecutive apastrons and its contribution to the accumulated angle is evidently $2 \pi w$. The number $z$ gives the number of ``leaves'' drawn by the solution before completing a full closed orbit. The path described by the geodesic will not necessarily follow consecutive ``leaves'' throughout the motion, so the number $v$ tells how many vertices the orbit will skip before drawing the next ``leaf''.

\section{The double sink solution}
Our second toy model for a BH binary is inspired in Analogue Gravity, a formal equivalence between the propagation
of sound waves in fluids and a scalar field in a (analogue) curved spacetime~\cite{Unruh:1980cg,Visser:1997ux,Barcelo:2005fc}. In particular,
a sound wave $\Phi$ associated with a velocity field $\vec{v}=\nabla\Phi$ is governed by the equation
\be
\Box\Phi=0\,.\label{KG}
\ee
The curved spacetime geometry associated with the D'Alembertian is fixed by the background fluid flow, and contains horizons if and when the 
fluid flow exceeds, at some point, the local sound speed~\cite{Unruh:1980cg,Visser:1997ux,Barcelo:2005fc}.
We consider a generalization of the $2+1$ ``draining bathtub'' geometry~\cite{Visser:1997ux}, which here describes two acoustic ``dumb'' holes. Our background flow is two-dimensional, consists on two sinks
of unit strength at $x=\pm a$, and can be written as~\cite{MilneBook},
\begin{subequations} \label{flowvel}
\begin{align} 
v_x&=-A\frac{2x(r^2-a^2)}{r^4 + 2a^2(-x^2+y^2)+a^4}\,,\\
v_y&=-Ay\left(\frac{1}{y^2+(x-a)^2}+\frac{1}{y^2+(x+a)^2}\right)\,,\\
v^2&=\frac{4A^2r^2}{r^4 +2a^2(-x^2+y^2)+a^4}\,,
\end{align}
\end{subequations}
with $r^2=x^2+y^2$. The constant $A > 0$ is arbitrary and fixes the fluid speed at some radius. For simplicity, we assume $A = 1$ throughout the paper.

\subsection{Horizons, ergoregions and soundcurves}
%
\begin{figure}[ht]
  \includegraphics[width=0.48\textwidth]{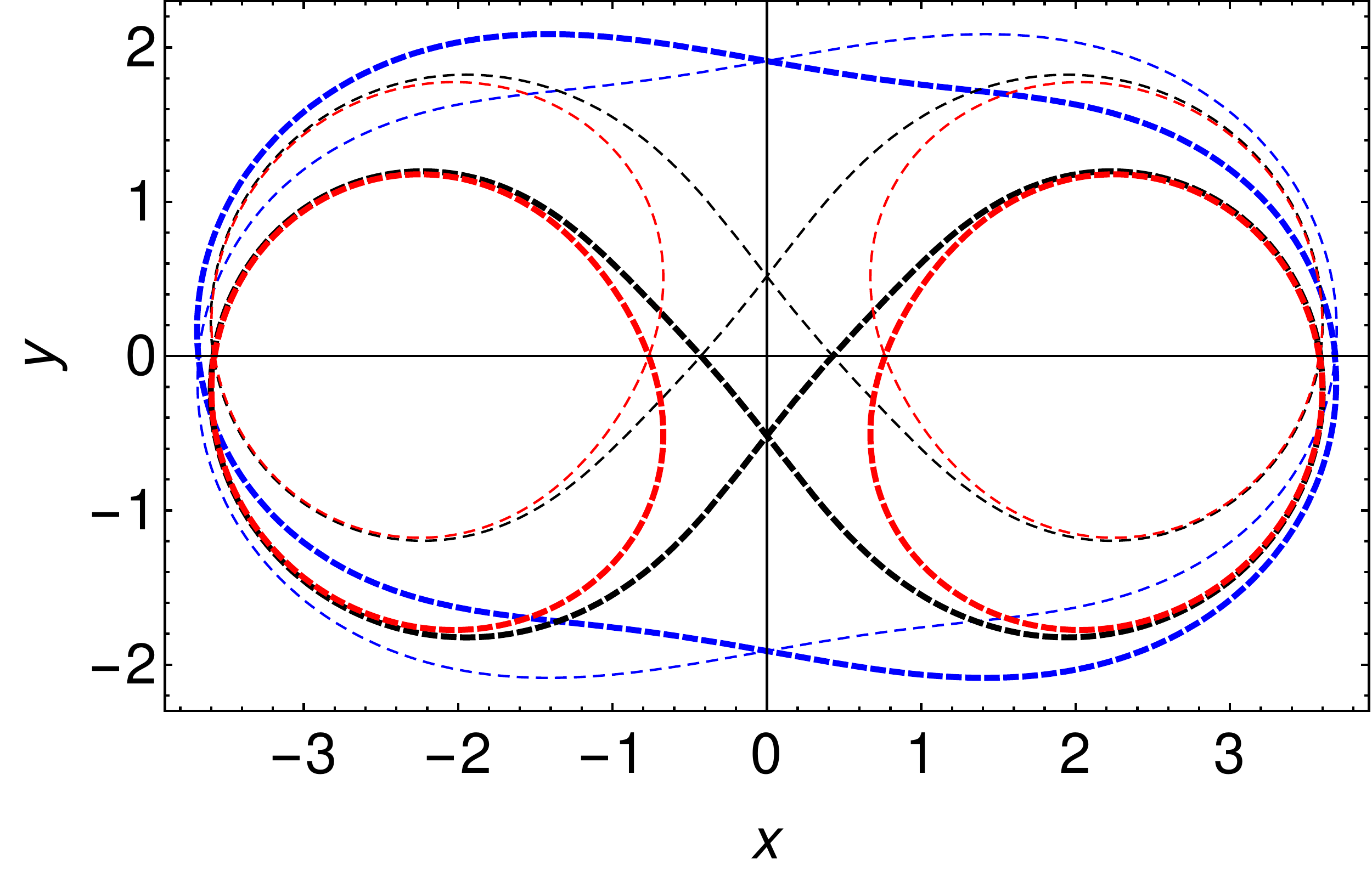}
  \caption{Different ``soundcurves'' when the separation between the sinks is $a=2$. These curves were computed by numerically solving the null geodesic equations for the analogue metric~\eqref{eq:sound-metric} with the velocity field~\eqref{flowvel}. \label{Soundcurve}}
\end{figure}
Some notions of curved spacetime geometry can be used to formally describe the propagation of sound waves, with velocity $c_s$, in irrotational fluids like the one determined by Eq.~\eqref{flowvel}. Among these, we find useful the following definitions. {\it Ergocurves} are curves at which $v^2=c_s^2$, and play the role of spacetime ergosurfaces. {\it Acoustic horizons}, on the other hand, are curves at which $\vec{n}\cdot\vec{v}=c_s$, where $\vec{n}$ is the normal vector at each point in the curve. Acoustic horizons play the role of event horizons when the acoustic geometry is stationary. Finally, {\it soundcurves} (which are analog light rings or analog photon surfaces) are closed curves for sound waves; a sound wave of velocity $v_s\equiv (dx/ds, dy/ds)$ (with $s$ an affine parameter) therefore satisfies, on a soundcurve,
$||\vec{v}_s-\vec{v}||=c_s$. These analog light rings were recently investigated experimentally in a rotating vortex representing the draining bathtub geometry~\cite{Torres:2017vaz}.

These properties can all be recovered using the effective metric experienced by sound waves (for simplicity we set the background density to unity),   
\beq
ds^2 &=&-(c_s^2-v^2)dt^2 - 2 \, dt (v_x dx+v_ydy)+dx^2+dy^2\,,\nonumber\\
&\equiv&- \left( \alpha^{2} - \beta^{i} \beta_{i} \right) \, dt^{2}
         + 2 \, dt \beta_{i} \,dx^{i}
         +   \gamma_{ij} \, dx^{i} \,dx^{j}\,.\label{eq:sound-metric}
\eeq
When there is no separation between the sinks ($a=0$), the metric above simplifies to
\be
ds^2 =-(c_s^2-4/r^2)dt^2 +\frac{4}{r} dt dr+dr^2+r^2d\phi^2\,.
\ee

Transforming $t$ and $r$ into new variables $\tilde t$ and $\tilde r$, defined by $r=\tilde{r}/c_s$ and $dt= d\tilde{t}/c_s^2+2\tilde{r}d\tilde{r}/\left(c_s^2(\tilde{r}^2-4)\right)$, brings the acoustic metric above to the canonical form~\cite{Berti:2004ju}
\be
c_s^2ds^2 =-\left(1-\frac{4}{\tilde{r}^2}\right)d\tilde{t}^2 +\frac{1}{1-\frac{4}{\tilde{r}^2}} d\tilde{r}^2+\tilde{r}^2d\phi^2\,.\label{acoustic_single}
\ee
%
\subsubsection{Soundcurves}
When the two holes are sufficiently close to one another, the metric reduces to \eqref{acoustic_single}, and the soundcurves are located at $r=2\sqrt{2}$~\cite{Cardoso:2008bp} (here and in all numerical calculations we assume $c_s=1$).
The coordinate time that sound takes to travel across this single ``merged sink'' is $T=8\pi$ (corresponding to a frequency $\Omega=1/2$), and the associated Lyapunov exponent is $\lambda=1/(2\sqrt{2})$.

For double sinks, we find that there are different sound curves, shown in Fig.~\ref{Soundcurve}. Global outer orbits (shown in blue) surrounding both BHs exist for any value of separation $a$. Inner orbits around each sink (shown in red) and ``8-shaped'' orbits (shown in black) exist only for values of separation $a>1$. It is worth noticing that these orbits are not symmetric, as opposed to the ones found in the Majumdar-Papapetrou spacetime. Furthermore, the global outer orbits can cross the other two types of orbits for small values of $a$, in contrast to what happens 
in MP spacetimes.

For double sinks separated by a large distance $a \gg 1$, the coordinate period of the global geodesic (encircling both holes) is $T\sim 15.6 \pm 0.3+4a$. This is in rough agreement with expectations: the total crossing time is of order $4a$ plus two semi-circles, each of which takes $2\pi$ to complete. In the same regime, the coordinate period of the inner closed lines is $\sim 12.5\sim 4\pi$,
as expected from the single-hole analysis. 

For separations $a\lesssim 5$ the travel time does not strongly depend on the separation itself, and is $\sim 8\pi$. Thus, depending on the dynamical aspect under investigation, the two sinks can be considered to be merged at these separations.

\subsubsection{Acoustic horizon}
%
\begin{figure}[ht]
  \includegraphics[width=0.48\textwidth]{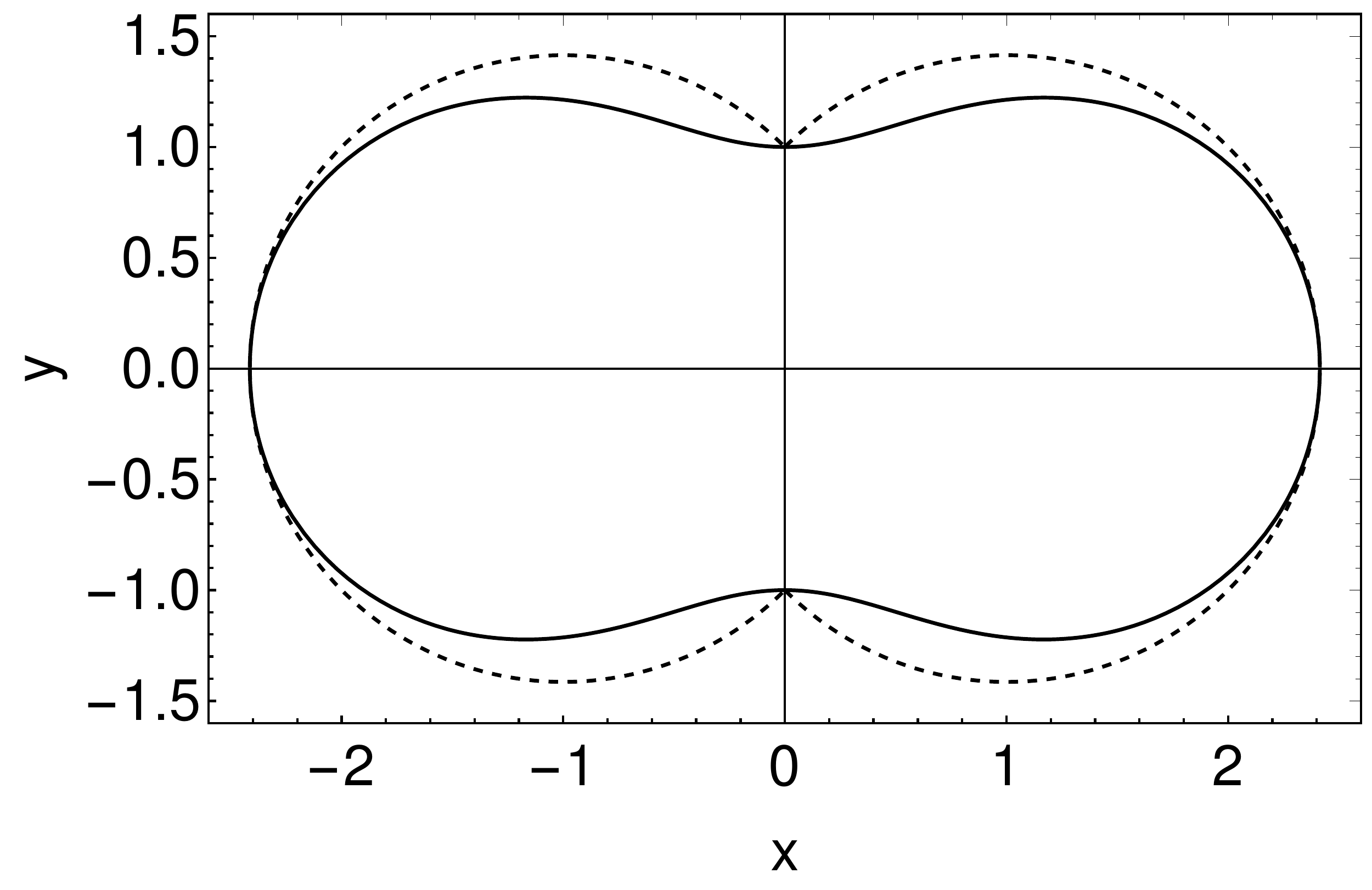}
  \caption{Acoustic horizon (solid line) surrounding both sinks for separation $a=1$. 
The horizon exists only if $a \leq 1$. The ergocurve (dashed line) clearly does not coincide with the horizon, except at a few points.
    \label{horizon}}
\end{figure}
Defining $\phi=y'(x)$, the normal to the curve $y(x)$ can be written as
\[
\vec{n}=\left(-\frac{\phi}{\sqrt{1+\phi^2}},\frac{1}{\sqrt{1+\phi^2}}\right)\,.
\]
It is straightforward to show that the equation $\vec{n}\cdot\vec{v}=c_s$ determining the horizon becomes a first order differential equation for $y(x)$:
\be
y' (x)+\frac{v_xv_y\pm \sqrt{c_s^2(v^2-c_s^2)}}{c_s^2-v_x^2}=0\,, \label{eq:hor_loc}
\ee
where the background velocities are given by \eqref{flowvel}. Since any acoustic horizon engulfing both sinks must necessarily intercept the symmetry line ($y$ axis), we look for solutions of \eqref{eq:hor_loc} that pass through $x=0$. We find that a common horizon engulfs both sinks only if $a  \leq 1$. When $a > 1$, the sinks are sufficiently apart and we recover a single horizon around each sink.

For $a  \le 1$, by searching the parameter space of the initial conditions of Eq.~\eqref{eq:hor_loc}, we have found that the acoustic horizon is always a smooth curve with $y(0)=\pm\left(1+\sqrt{1-a^2}\right)$ and $y'(0)=0$. In Fig.~\ref{horizon} we plot the shape of the horizon at the critical separation $a=1$.

\subsubsection{Ergoregion}
Although a notion and definition of ergoregion for dynamical
spacetimes does not exist, the acoustic version described above is readily extended to any geometry. Solving the equation $v^2 =c_s^2$ for the double sink flow given by Eq.~\eqref{flowvel} yields
\be
y = \pm\frac{1}{c_s}\sqrt{2-a^2 c_s^2 - x^2 c_s^2 + 2 \sqrt{1+a^2 c_s^2 (x^2 c_s^2 -1)}} \,.
\ee
The function $y(x)$ informs on the location of the ergocurve. In general, the acoustic horizon and the ergocurve do not coincide. To show this, assume that they do and that $v^2 = c_s^2$ everywhere along the horizon surface. At any given point $x$ where $y(x)$ is a smooth function, and $v_x$ and $v_y$ are non-zero, the horizon equation \eqref{eq:hor_loc} implies that
\be
\frac{dy}{dx} + \frac{v_x}{v_y} = 0 \Rightarrow \frac{v_y}{v_x} + \frac{v_x}{v_y} = 0 \Rightarrow v^2 = 0\,,
\ee
which clearly contradicts the assumption that $v_x\neq 0$ and $v_y \neq 0$.  Consequently, the ergocurve and the acoustic horizon can only coincide at points for which either $v_x=0$ or $v_y=0$ or $y(x)$ is not smooth. This fact is illustrated in Fig.~\ref{horizon}, which exhibits four points where they coincide.

In other words, for acoustic binaries the horizon lies {\it inside} the ergocurve, raising the interesting prospect of phenomena such as superradiance~\cite{Brito:2015oca,Torres:2016iee} to occur. These results suggest that similar phenomena may be present in gravitational BH binaries which may lead to new effects during the inspiral.

\subsection{Wave scattering and quasinormal modes}
There is an established connection between photon surfaces and the dynamical response of perturbed, single-BH spacetimes~\cite{Cardoso:2008bp,Cardoso:2017cqb,Cardoso:2017njb,Chaverra:2015aya}.
In a nutshell, the photon surface works as a trapping region where scalar, electromagnetic or gravitational perturbations
can linger. The instability timescale of null geodesics is then related directly to the lifetime of massless perturbations. 

We wish to understand if such connection carries over to binaries.
The analysis of wave dynamics in BH binaries is notoriously difficult. For single BH spacetimes with certain symmetries, a linearized study where the relevant partial differential equations are transformed into ordinary differential equations is possible.
Sound waves $\Phi(t,r,\phi)\sim \Phi(t,r)e^{im\phi}$ in single acoustic dumb-hole spacetimes were studied years ago, and it was concluded that~\cite{Berti:2004ju,Cardoso:2004fi}:
\begin{enumerate}[{\bf i.}]

\item The early-time response is strongly dependent on the initial conditions, but is followed by quasinormal ringdown~\cite{Berti:2009kk}.
In the eikonal regime, such ringdown is described by sound waves circling the soundcurve and decaying away. For the geometry \eqref{acoustic_single}, the fundamental (longest-lived) frequencies $\omega_m$ for the first few azimuthal numbers $m$ are~\cite{Berti:2004ju,Cardoso:2004fi}\footnote{These numbers were obtained using a continued fraction representation~\cite{Berti:2009kk}.}(see \cite{Patrick:2018orp} for a ringdown analysis in presence of vorticity):
\begin{subequations}
\begin{align}
%
\omega_1&=0.20342-0.17062i\,,\\
\omega_2&=0.47636-0.17537i\,,\label{acoustic:dipole}\\
\omega_3&=0.73427-0.17621i\,,\\
\omega_4&=0.98823-0.17648i\,.
\end{align}
\end{subequations}

\item The ringdown stage eventually gives way to a late-time power law tail, where the field decays as
\be
\Phi\sim t^{-(2m+1)}\,.\label{power-law}
\ee
\end{enumerate}

Appealing to the correspondence between quasinormal modes and geodesic properties~\cite{Cardoso:2008bp,Dolan:2011ti}, one expects a relation of the type
\be
\omega_m=m\Omega-i(n+1/2)\lambda=\frac{m}{2}-i\frac{(2n+1)}{4\sqrt{2}}\,,\label{geo_qnms}
\ee
to hold at large $m$. This relation indeed describes extremely well the results above for the quasinormal modes of acoustic holes\footnote{In fact, one can immediately see that the imaginary part is captured to better than $4\%$ already for the $m=1$ mode.}. Nevertheless, as shown in Ref.~\cite{Dolan:2011ti} for the draining bathtub geometry, expansions of $\omega_m$ can be obtained to arbitrary order in powers of $m$. 

\subsubsection{Numerical approach}
For the numerical evolutions of \eqref{KG}, we used the \textsc{EinsteinToolkit} infrastructure~\cite{Loffler:2011ay,EinsteinToolkit:web,Zilhao:2013hia}, which is based on the \textsc{Cactus} Computational Toolkit~\cite{Cactuscode:web}, a software framework for high-performance computing.
The spacetime metric is fixed according to Eq.~(\ref{eq:sound-metric}), and we evolve the Klein-Gordon equation on top of this background using the same numerical code as the one employed in~\cite{Cunha:2017wao}, herein accordingly adapted to 2+1 evolutions. To reduce the system to a first order form, we introduce the ``canonical momentum'' of the scalar field $\Phi$
\begin{equation}
\label{eq:Kphi}
K_{\Phi} = -\frac{1}{2\alpha}  \left( \partial_{t} - \Lie_{\beta} \right) \Phi \,,
\end{equation}
where $\Lie$ denotes the Lie derivative, to write our evolution system in the form
\beq
\partial_{t} \Phi &=& - 2 \alpha K_\Phi + \Lie_{\beta} \Phi \label{eq:dtPhi} \nonumber\\
\partial_{t} K_\Phi & =& \alpha \left( K K_{\Phi} - \frac{\gamma^{ij}}{2}  D_i \partial_j \Phi\right)
%
- \frac{\gamma^{ij}}{2}  \partial_i \alpha \partial_j \Phi+ \Lie_{\beta} K_\Phi \,, \label{eq:dtKphi}\nonumber
\eeq
where $D_i$ is the covariant derivative with respect to the $2$-metric
$\gamma_{ij}$ and $K$ is the trace of the extrinsic curvature
$K_{ij} = \frac{1}{2\alpha} \Lie_{\beta} \gamma_{ij}$.

To numerically evolve this system, we employ the method-of-lines, where spatial
derivatives are approximated by fourth-order finite difference stencils, and we
use the fourth-order Runge-Kutta scheme for the time integration. Kreiss-Oliger
dissipation is applied to evolved quantities in order to damp high-frequency
noise. A complication arising from metric~(\ref{eq:sound-metric}) is the presence of curvature singularities at $x = \pm a$. To deal with these we use a simple excision procedure, whereby our evolved quantities are multiplied by a mask function $M(x,y)$
\[
  \partial_t u \rightarrow M(x,y) \, \partial_t u \,,
\]
where $u=\Phi,K_{\Phi}$ and $M(x,y)$ is a smooth transition function that evaluates to 1 outside the horizon and 0 inside.\footnote{We thank D.~Hilditch for discussions about this point.}

We evolve the system with initial conditions given by
\begin{subequations}
\begin{align}
\Phi(x,y)&= \left( c_0 + c_1 \frac{X}{R} + c_2 \frac{X^2 - Y^2}{R^2}
\right) e^{-\frac{(R-R_0)^2}{2 \sigma^2}} \,, \\
K_{\Phi}&= 0 \,,
\end{align} \label{eq:ID}
\end{subequations}
where $X=x-x_0$, $Y=y-y_0$ and $R=\sqrt{X^2+Y^2}$. The constants $c_m$ characterize the multipole of order $m$. The function $\Phi(x,y)$ represents a Gaussian pulse of width $\sigma$ centered at $(x_0,y_0)$ and peaked at $R=R_0$.

\subsubsection{Results}
%
\begin{figure}[htbp]
  \centering
  \includegraphics[width=0.5\textwidth]{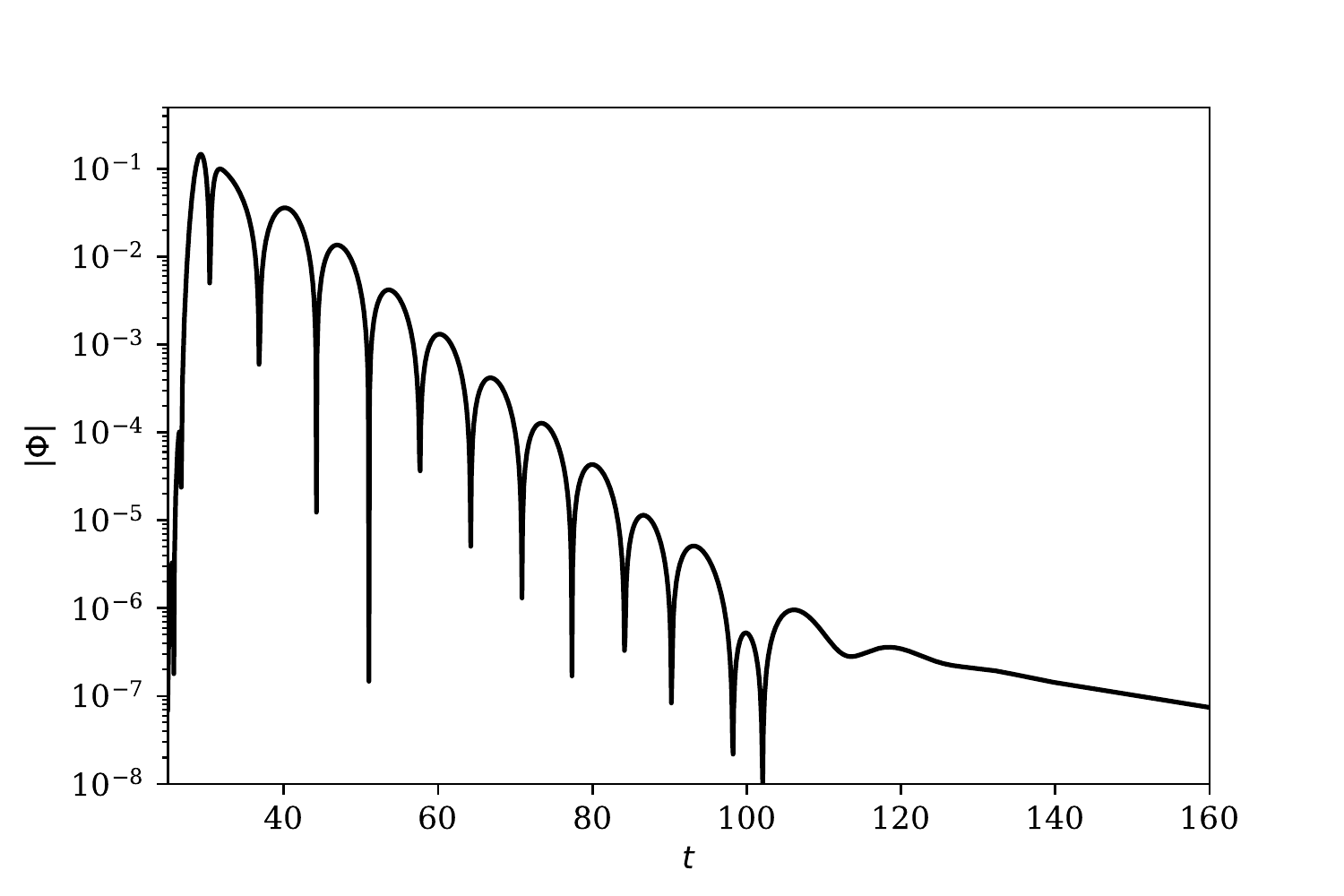}
  \caption[]{\label{fig:single_m2_phi} Evolution of a quadrupole ($c_0=c_1=0, c_2=1$) gaussian (with $x_0=y_0=0$, $R_0=5, \sigma=1/2$) profile around a single ($a=0$) acoustic hole. 
	The scalar was extracted at $x=30$, $y=0$. A ringdown, exponentially decaying, phase is visible, and followed by a late-time power-law tail.
	This and subsequent plots show the absolute value of the scalar.\label{fig:single_evol} }
\end{figure}
We extract the scalar field $\Phi$ at $x=30, y=0$ and at $x=0, y=30$. The first point lies along the symmetry line; the second extraction point
lies along the line joining the two sinks, and is closer to one than to the other. 
The general features of our results remain the same at other extraction points.

Figure \ref{fig:single_evol} shows the result of evolving a Gaussian, quadrupolar initial data, in the background of a single sink ($a=0$). The signal is exponentially damped at intermediate times, and we find excellent agreement (to within $\sim 2\%$) with prediction \eqref{acoustic:dipole}, and hence also with the soundcurve analysis. The late-time tail, also clear in Fig.~\ref{fig:single_evol}, is well described by $\Phi\sim t^{-4.9}$, in good agreement with prediction~\eqref{power-law}.

\begin{figure}[ht]
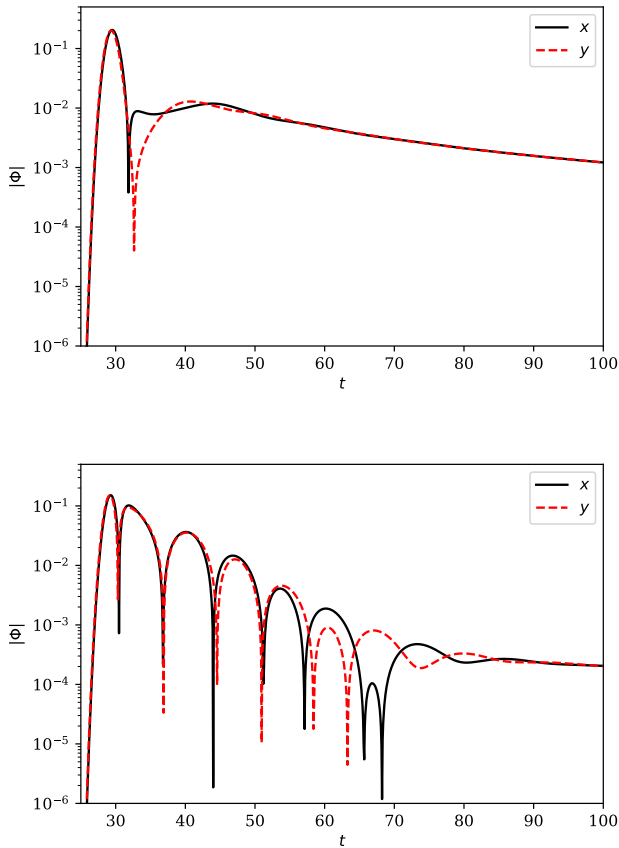

\begin{tabular}{c}
\includegraphics[width=0.5\textwidth,clip]{{{phi_x0y0_double_m0_r0_5_sigma_0.5_a_1}}}\\
\includegraphics[width=0.5\textwidth,clip]{{{phi_x0y0_double_m2_r0_5_sigma_0.5_a_1}}}
\end{tabular}
\caption{Scattering of a sound wave off a binary dumb-hole with $a=1$. The Gaussian, centered at the origin, has a width $\sigma=1/2$ and is localized at $R_0=5$. 
The field $\Phi$ is extracted along the $x$-axis (black solid line) and along the $y$-axis (red, dashed line) at a distance of $30$ from the origin.
{\bf Top:} Monopolar wave, with $c_0=1$. For monopolar waves the prompt signal quickly gives way to a power-law tail $\Phi \sim t^{-1.1}$ at late times, in agreement with prediction~\eqref{power-law}.
{\bf Bottom:} Quadrupolar wave, with $c_2=1$. We find a ringdown $\omega=0.472-0.170i$ and a power law $\sim t^{-1.2}$.
}
\label{fig:acoustic_small_separation}
\end{figure}
Figure~\ref{fig:acoustic_small_separation}
shows the result of evolving monopolar and quadrupolar waves in the background of a binary with finite but small separation, $a=1$. We find that the signal still has all the features of the single-BH spacetime: a single ringdown stage, followed by a late-time power law tail. Both the ringdown and the tail are well-described by single BH geometries.
In particular, due to mode-mixing caused by the lack of axisymmetry, an initial pure-$m$ data will evolve to a superposition of different $m$'s. In particular,
the $m=0$ mode seems to be present at late times in the waveform and causes the latetime power-law tail to be well described by $\Phi \sim 1/t$ (cf. Eq.~\eqref{power-law}).
In other words, binaries separated by a distance $a\lesssim 1$ behave very similarly to a single BH. This is not completely surprising and is most likely connected
to such binaries having a common horizon (cf. Fig.~\ref{horizon}). This feature is also present in nonlinear evolutions of Einstein equations for the BH-binary problem,
where ringdown is seen to be triggered as soon as merger occurs, and in a very linear way~\cite{Buonanno:2006ui,Berti:2007fi,London:2018gaq,Baibhav:2017jhs,Brito:2018rfr}.

\begin{figure}[ht]
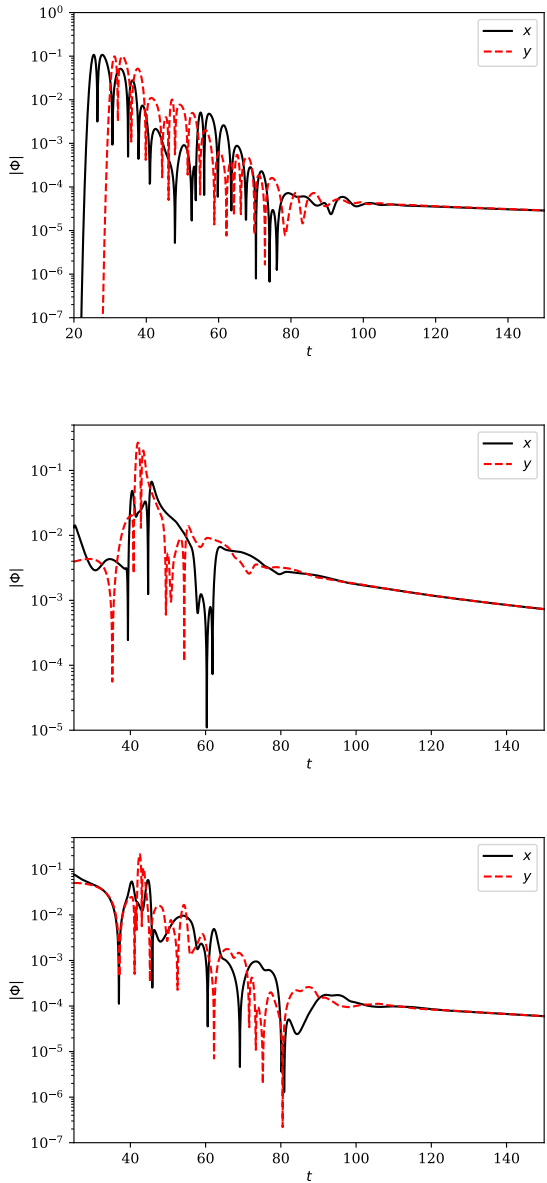

\begin{tabular}{c}
\includegraphics[width=0.45\textwidth,clip]{{{phi_x0y0_double_m2_x0_5_r0_3_sigma_0.5_a_5}}}\\
\includegraphics[width=0.45\textwidth,clip]{{{phi_x0y0_double_m0_r0_10_sigma_0.5_a_5}}}\\
\includegraphics[width=0.45\textwidth,clip]{{{phi_x0y0_double_m2_r0_10_sigma_0.5_a_5}}}
\end{tabular}
\caption{Scattering of a sound wave off a binary dumb-hole with $a=5$. The gaussian width is $\sigma=1/2$.
The field $\Phi$ is extracted along the $x$-axis (black solid line) and along the $y$-axis (red, dashed line) at a distance of $30$ from the origin.
{\bf Top:} Quadrupolar wave, with $c_2=1$, $x_0=5$, $y_0=0$, $R_0=3$. First ringdown stage for $x$ matches well that of a one single, individual BH. Second ringdown is of lower frequency. Ratio is 1.2.
{\bf Middle:} Monopolar wave, with $c_0=1$, $x_0=y_0=0$, $R_0=10$.
{\bf Bottom:} Quadrupolar wave, with $c_2=1$, $x_0=y_0=0$, $R_0=10$.}
\label{fig:large}
\end{figure}
When the separation is made to increase, the different scales in the problem become noticeable, as seen in Fig.~\ref{fig:large}. For $a=5$, the coordinate period of the global geodesic is of order $\sim 35$, larger by about a factor three than the period of the geodesics encircling each BH.

Focus on the top panel of Fig.~\ref{fig:large}. The initial data is localized around only one of the BHs (located at $x=5$). Therefore the initial data excites this BH first
and to a higher extent than the second BH, farther apart. On the $x$-axis, the initial ringdown is well described by the modes of this BH. After a time $\sim 25$ the ringdown mode of the perturbed rightmost BH reaches the extraction point $x=30$ on the $x$-axis (black solid curve). Later, starting at $t \sim 55$ the ringdown of the leftmost BH can be seen, followed by the decaying power-law $\Phi\sim t^{-1}$.
The global mode is encoded in the envelope of this signal, and the only reason it is not more clearly visible
is because the power-law tail sets in very quickly. Such behavior is similar to the setting in of global modes of anti-de Sitter spacetime, or of ultracompact exotic objects~\cite{Cardoso:2015fga,Cardoso:2017cqb,Cardoso:2016rao,Correia:2018apm}.

The two lower panels in Fig.~\ref{fig:large} show the result of evolving a monopolar and quadrupolar data, centered at the origin and localized away from the two holes.
The signal is peaked at some intermediate time, most likely the result of interference between different modes. We do not have a simple and elegant explanation for such feature.
For any initial data we looked at, late-time tail is always dominated by $t^{-1}$, that of the circularly symmetric mode.

\section{On the structure of circular null geodesics for eternal and for coalescing binaries}
%
\begin{figure*}[t]
\begin{center}
\begin{tabular}{ccc}
\includegraphics[width=0.33\textwidth]{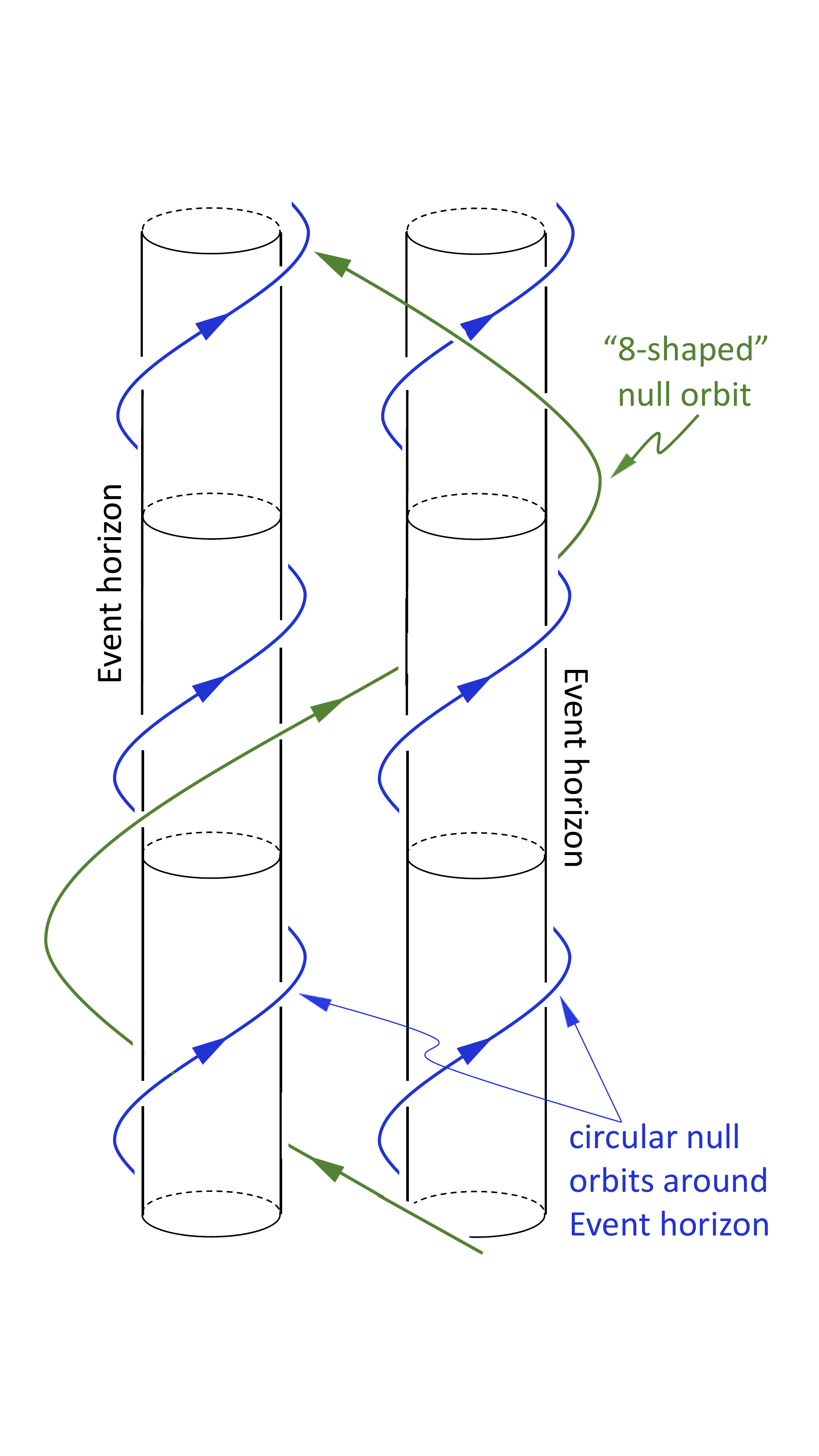} &
\includegraphics[width=0.33\textwidth]{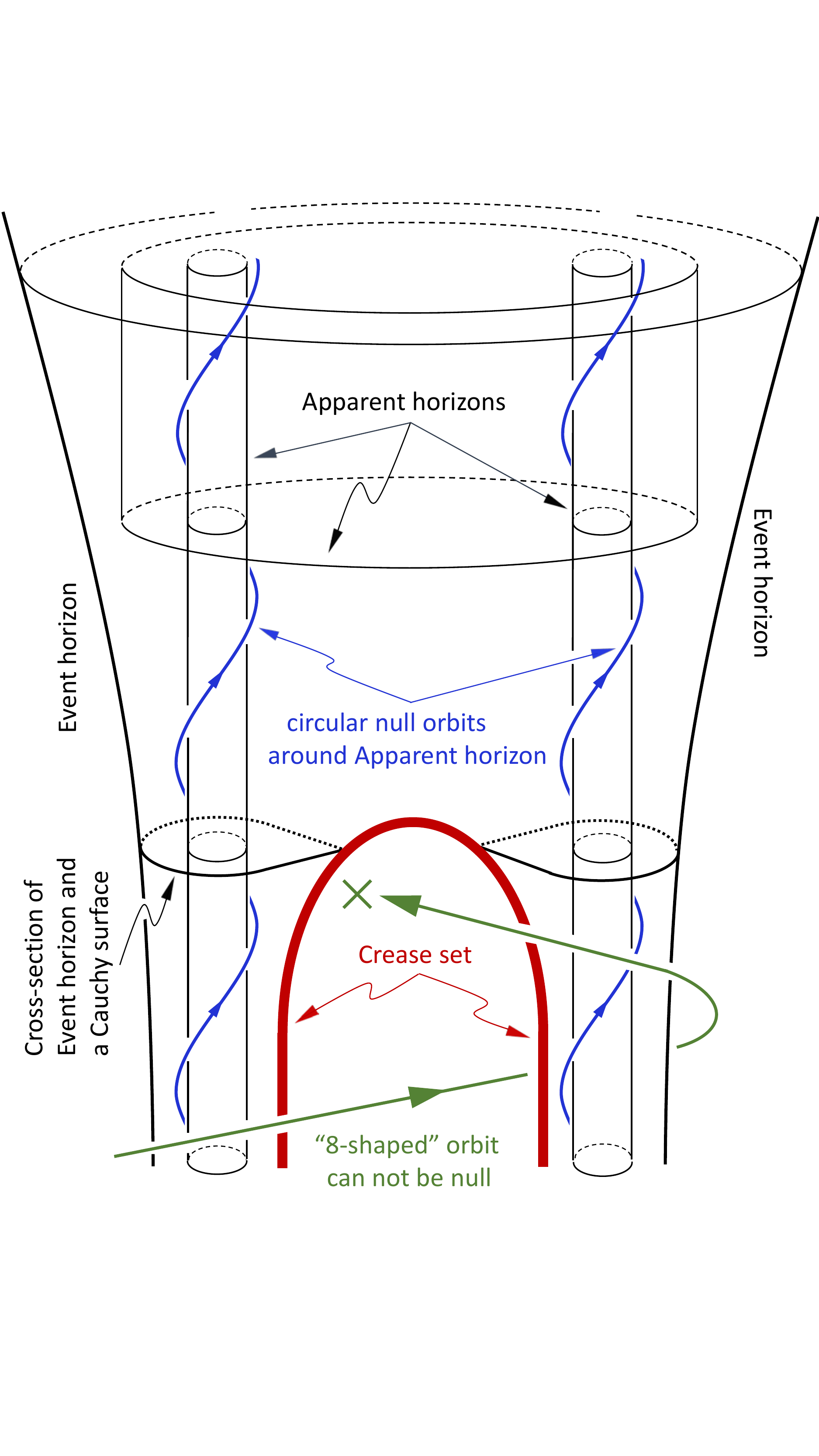}&
\includegraphics[width=0.33\textwidth]{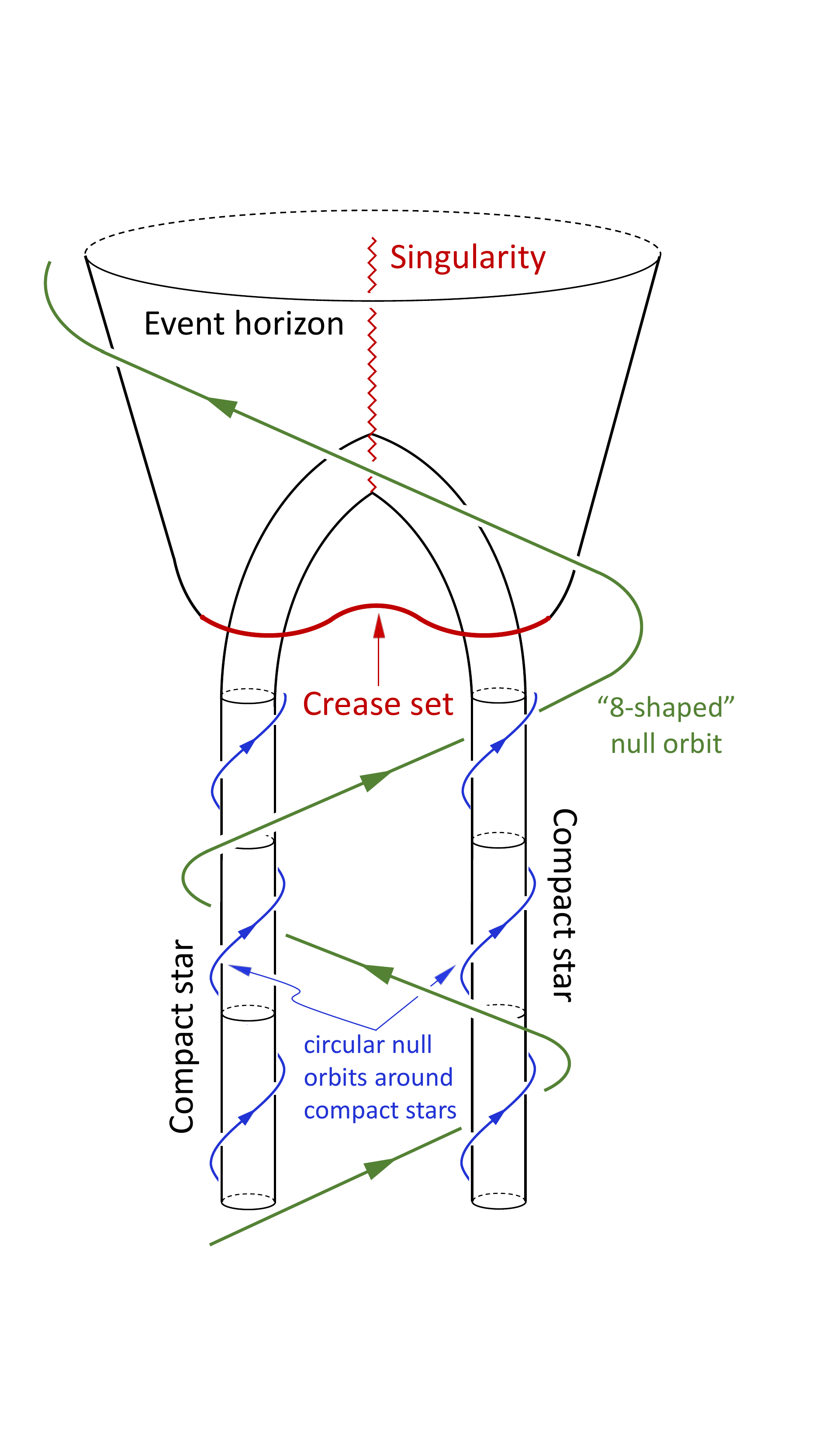}
\end{tabular}
\caption{{\bf Left:} Null geodesic curves in two MP BHs, whose event horizon consists of two disconnected components. 
  Depicted are two circular orbits around each of the disconnected components (blue lines) and a ``8-shaped'' orbit (green line) around the event horizon.  
		{\bf Middle:} Two BHs merge to form a single BH. The event horizon has only a single connected component and admits a ``crease set'' (depicted by thick red-line) consisting of the past end points of some of the null geodesic generators of the event horizon. Here ``two black holes'' is meant to be the number of cross-sections of 
  the event horizon on a particularly chosen Cauchy time slice as illustrated. Note however that since the event horizon has, as a part, the crease set, which is spacelike, 
  the notion of the number of cross-sections (as well as the topology of the cross-section) is dependent on the choice of Cauchy time slice. 
  Also because of the spacelike nature of the crease-set, null/timelike orbits cannot be ``8-shaped.''  
  Inside the event horizon there are apparent horizons, for which null geodesic curves can possibly form circular orbits (blue lines). 
{\bf Right:} Two ultra-compact stars collide to form a single BH. The system can admit a ``8-shaped'' null orbit (green line) as well as circular null orbits (blue lines) 
  around each of the ultra-compact stars if they are sufficiently compact.  Further study is needed to see whether these ``8-shaped'' and circular null orbits eventually 
  can go around outside the event horizon (as illustrated here) or must go inside the event horizon, after the BH formation. 
\label{fig:pants} 
}
\end{center}
\end{figure*}
One may think that even though our geometries of BH binaries described by the MP metric are stationary, they can be interpreted as a “snapshot” of a dynamically coalescing process of two BHs, and therefore the sequences of the circular null geodesics obtained in this paper could mimic a photon surface around a dynamical geometry of colliding BHs in a more astrophysically realistic situation. However, there is a sharp contrast between the stationary MP binary and a dynamically colliding BH binary. One should first note that the BH binaries in the MP solution never merge to form a single BH; they have two mutually disconnected components of the event horizon and therefore allow for closed null/timelike orbits outside the event horizon (see left panel of Fig.~\ref{fig:pants}). In contrast, when two BHs dynamically coalesce, the event horizon as a three-dimensional null hypersurface has only a single connected component from the beginning.

Let us recall that the future event horizon of a BH is formed by null geodesic generators which have no end points in the future but can in general admit past end points. For stationary BHs, the event horizon is a Killing horizon and the horizon null geodesic generators have no past end points.
As for dynamical BH system, such as those formed by gravitational collapse or by the merger of two BHs, (some of) the horizon null geodesic generators have {\em past} end points, which form a subset in the spacetime. An important feature is that such a past-end-point set--called {\em crease set}--must be {\em spacelike} 
and {\em achronal} rather than timelike or null, since the event horizon itself must be an achronal null hypersurface. Now when we say, “a single” BH or “two” BHs, we refer to as the number of the cross-section components of the event horizon and a Cauchy surface. It has been shown in \cite{Siino:1997ix} that this notion of the “number of cross-sections” is in fact time-slice dependent when the event horizon admits a crease set as a part, since in that case one can choose a Cauchy slice such that it crosses the crease-set as many times as one wishes (middle panel of Fig.~\ref{fig:pants}). 
Furthermore, depending upon the choice of Cauchy slicing, even the topology of the horizon cross-section changes (for instance, when the crease set is 2-dimensional, it can even be temporarily toroidal~\cite{Hughes:1994ea,Shapiro:1995rr,Emparan:2017vyp}). For the mathematical structure of the crease set and the classification of the possible (temporal) horizon cross-section topologies, see Ref.~\cite{Siino:1997ix}. 
For the dynamically coalescing “two” BH binary, the temporarily “two BHs” are connected by a crease set, and the achronal nature of this crease set prevents any causal geodesic curve (either null or timelike) from forming a closed orbit around either ``each single'' BH or the ``8-shaped'' closed orbit that would encompass the “two BHs.” An exception is a global outer closed orbit. Therefore one has to be careful when attempting to make a physical interpretation of closed null orbits found in static MP solution.

The observation above would be manifest if, for example, one examines closed null geodesics in Kastor-Traschen (KT) solution~\cite{Kastor:1992nn} with two BHs. The KT solution also describes configuration of multiple BHs with maximal charge, i.e., $M=|Q|$, but it includes a positive cosmological constant. Therefore those multiple BHs are on a de Sitter expanding universe. The time-reverse of the solution can describe coalescing BHs, and their event horizon must have crease sets. Since each BH has its own apparent horizon inside the event horizon, one can presumably find e.g., ``8-shaped'' closed null orbits that go around two “apparent horizons” but the orbits should entirely be inside the event horizon.  

If we consider, instead of a coalescing BH binary, a coalescing compact star binary, each of the stars is compact enough to admit its own photon sphere, then we can find ``8-shaped'' closed null/timelike orbits, besides the global null closed orbits. See right-most panel of Fig.~\ref{fig:pants}.

The above notes of caution refer to the purely formal description of such geodesics and event horizons. Our detectors are placed some finite distance away from the black hole binary and are active for a short amount of time: they have no access to a teleological quantity such as eternal horizons. Instead, they measure properties related to local quantities, such as apparent horizons.
Therefore, the geodesics that we describe should capture basically the physics at play in gravitational-wave detectors.

\section{Conclusion}
We have given the first steps in our understanding of wave dynamics around BH binaries. We found three types of null geodesics,
which should play a role in the decay of perturbed binaries. Our results for sound waves in an acoustic geometry indeed indicate
that the different timescales can be associated to such geodesics. There are many open issues, concerning the specific features of the wave dynamics, and how it generalizes when the binary itself has dynamics. In particular, defining and exploring the notion of closed null orbits and ergoregions for dynamical spacetimes is challenging, but potentially important, as it raises the possibility of gravitational lensing effects, superradiant effects and Penrose energy extraction.

\begin{acknowledgments}
T.~A.\ is grateful to Instituto Superior T\'ecnico for hospitality where part of this work was carried out. 
T.~A.\ acknowledges support from the S\~ao Paulo Research Foundation (FAPESP), Grants No.\ 16/00384-1 and 17/10641-4. 
V.~C.\ is indebted to Kindai University in Osaka for hospitality where this work was started.
V.~C.\ acknowledges financial support provided under the European Union's H2020 ERC Consolidator Grant ``Matter and strong-field gravity: New frontiers in Einstein's theory'' grant agreement no.\ MaGRaTh--646597.
M.~R.\ acknowledges partial financial support from the S\~ao Paulo Research Foundation (FAPESP), Grant No. 2013/09357-9, and from Conselho Nacional de Desenvolvimento
Cient\'ifico e Tecnol\'ogico (CNPq).
M.~Z.\ is funded through Program No. IF/00729/2015 from FCT (Portugal) and would also like to thank the hospitality of Kindai University in Osaka where part of this work was done.
This work has received funding from the European Union's Horizon
2020 research and innovation programme under the Marie Sk\l odowska-Curie
grant agreement No 690904.
This work was supported in part (A.~I.) by JSPS KAKENHI Grant Number 15K05092, 26400280. 
The authors would like to acknowledge networking support by the COST Action CA16104.
\end{acknowledgments}

\appendix 
\section{Geodesics in the Majumdar-Papapetrou geometry: the symmetry plane\label{app:geoMP}}
This appendix provides further details on the geodesics along the symmetry plane of the MP geometry.

\subsection{Integration of geodesic equations in the MP metric\label{app:integration_geo}}
The geodesic equations are found by varying the action with respect to the Lagrangian in Eq.~\eqref{eqn:lagrangian_MP}. The equations of motion are the well know Euler-Lagrange equations:
\be
\frac{d}{d \lambda} \frac{\p \mathcal{L}}{\p \dot{x^{\mu}}} - \frac{\p \mathcal{L}}{\p x^{\mu}} = 0,
\ee
where $x^{\mu} = \{t,\rho, \phi, z\}$, and $\lambda$ is an affine parameter. The two conserved quantities are $E=\dot{t}/U^2$ and $L=\rho^2 U^2 \dot{\phi}$, which we identify as energy and angular momentum.

 We can find an energy conservation equation by substituting $E$ and $L$ into Eq.~\eqref{eqn:lagrangian_MP}:
%
\begin{align}
\label{eqn:energy_conservation_MP1}
\dot{\rho}^2 + \dot{z}^2 + V_{\rm eff}(\rho, z)  = E^2, \\
\label{eqn:effective_potential_MP}
V_{\rm eff}(\rho, z) = \frac{L^2}{\rho^2 U^4} + \frac{\delta}{U^2}.
\end{align}
Here and throughout the appendix, a dot means derivative with respect to the affine parameter $\lambda$. 
In terms of $E$ and $L$, the Euler-Lagrange equations reduce to the following system of coupled differential equations for $\rho$ and $z$:
\begin{align}
&\ddot{\rho} - \frac{L^2(U + \rho \p_{\rho}U)}{\rho^3 U^5} + \frac{2 \dot{\rho} \dot{z} \p_z U - (E^2+ \dot{z}^2 - \dot{\rho}^2) \p_{\rho} U}{U}=0, \\
&\ddot{z} - \frac{L^2 \p_z U}{\rho^2 U^5} + \frac{2 \dot{\rho} \dot{z} \p_{\rho} U - (E^2 - \dot{z}^2 + \dot{\rho}^2) \p_z U}{U} =0.
\end{align}
To find symmetric orbits we set initial conditions as $\rho(0)=0$, $z(0)=z_0$, and $\dot{z}(0)=0$. By virtue of Eqs.~\eqref{eqn:energy_conservation_MP1} and~\eqref{eqn:effective_potential_MP} we find
\be
\dot{\rho}(0) = \pm \sqrt{E^2 -V_{\rm eff}(0,z_0) }.
\ee
With the above initial conditions, we are able to find various orbits by choosing values of $E$, $L$, and $z_0$. Setting $L=0$ means constraining the motion to the plane that contains both holes, as explained in the main text. On the other hand, to look at the symmetry plane $z=0$ we need to choose non-zero values for $\rho(0)$ and $L$.

\subsection{Oscillating particle in the MP metric}
We start with a special case of the three-body problem in which one the bodies is much lighter than the other two. The small body is a test particle following a timelike geodesic motion at $z=0$, while the two massive bodies are the two extremely charged BHs in the Majumdar-Papapetrou metric. Setting $\delta=1$ in Eq.~\eqref{eqn:lagrangian_MP} yields
\be
\dot{\rho}^2 + \frac{1}{U^2}= E^2\,,\label{eqn:radial_azimuth}
\ee
where $U=U(\rho,z=0)$.
The effective potential $1/U^2$ tends to $1$ as $\rho \rightarrow \pm \infty$ and $z=0$. Thus, the solution is bounded and oscillatory around $\rho=0$ for $1/(1+2M)^2<E^2<1$. The first inequality comes from the fact that the initial velocity must be real. The motion can be parametrized by the coordinate time $t$ by dividing both sides of Eq.~\eqref{eqn:radial_azimuth} by $\dot{t}^2$:
\begin{equation}
\label{eqn:radial_azimuth_coord_time}
\left( \frac{\dot{\rho}}{\dot{t}} \right)^2 =\left(\frac{d \rho}{dt}\right)^2 = \frac{1}{E^2 U^4}\left( E^2 -\frac{1}{U^2} \right).
\end{equation}

For small-amplitude oscillations, 
\be
V\equiv \frac{1}{U^2}\sim \frac{1}{(1+2M)^2}+\frac{2M\rho^2}{(1+2M)^3}+{\cal O}(\rho^3)\,.
\ee
Thus, to lowest order $E^2=1/(1+2M)^2$ and the motion is sinusoidal with (coordinate) frequency
\be
\omega=\sqrt{\frac{2MV_0}{(1+2M)^5}}\,,
\ee
with $V_0=1/(1+2M)^2$.
For large separations (in our units, when $M\to 0$) we recover the Newtonian result.

\subsection{Lyapunov exponent\label{app:Lyapunov}}
One can compute the Lyapunov exponent associated with the unstable circular orbits by considering perturbations in the radial coordinate $\rho$ and in the canonical momentum $p_{\rho} = \p \mathcal{L} / \p \dot{\rho}$~\cite{Cornish:2003,Cardoso:2008bp}. Let $p_{\rho}=\delta p_{\rho}$ and $\rho = \rho_{\rm out} + \delta \rho$, then the equation for the perturbation reads
\begin{equation}
    \label{eqn:perturbation_matrix}
    \frac{d}{dt}
    \left(
        \begin{matrix}
            \delta p_{\rho} \\
            \delta \rho
        \end{matrix}
    \right)
    =
    \left(
    \begin{matrix}
        0 & K_1 \\
        K_2 & 0 \\
    \end{matrix}
    \right)
    \left(
        \begin{matrix}
            \delta p_{\rho} \\
            \delta \rho
        \end{matrix}
    \right)\,,
\end{equation}
where 
\be
K_1=\dot{t}^{-1} \frac{d}{d \rho}\left( \frac{\p \mathcal{L}}{\p \rho} \right)\,,\quad K_2=(\dot{t}U^2)^{-1}\,. \label{eqn:K}
\ee
The Lyapunov exponent $\lambda$ is the eigenvalue of the matrix in Eq.~\eqref{eqn:perturbation_matrix}, namely $\sqrt{K_1 K_2}$. It is worth noting that the Lyapunov exponent depends on the choice of parametrization. For that reason the perturbed equation is written as an ordinary differential equation with respect to the coordinate time $t$, so that associated time scale is the one measured by an observer at infinity. By virtue of the Euler-Lagrange equations, 
\begin{equation}
    \label{eqn:euler-lagrange}
    \frac{\p \mathcal{L}}{\p r} = \frac{d}{d \lambda}(g_{\rho \rho}\dot{\rho}) = 
    \dot{\rho} \frac{d}{d \rho}(g_{\rho \rho}\dot{\rho})\,.
\end{equation}
Taking the derivative of Eq.~\eqref{eqn:null_radial} with respect to $\rho$ yields
\begin{equation}
\label{eqn:null_radial_derivative}
\dot{\rho}\frac{d \dot{\rho}}{d \rho} = -\frac{1}{2}V'\,,
\end{equation}
where $V'=dV/d\rho$. From Eqs.~\eqref{eqn:null_radial} and~\eqref{eqn:null_radial_derivative},
\beq
\dot{\rho} \frac{d}{d \rho}(g_{\rho \rho}\dot{\rho}) &=& g_{\rho \rho}\dot{\rho} \frac{d \dot{\rho}}{d \rho} + \dot{\rho}^2 g'_{\rho \rho} =-\frac{1}{2}g_{\rho \rho} V' + (E^2-V)g'_{\rho \rho} \nonumber \\
%
&=&\frac{1}{2 g_{\rho\rho}} [g^2_{\rho \rho}(E^2 -V)]'\,.\label{eqn:enull1}
\eeq

For circular geodesics $\dot{\rho}=0$, which implies that $V=E^2$ and that the radius is a critical point of the potential.
Eqs.~\eqref{eqn:euler-lagrange} and~\eqref{eqn:enull1} imply
\begin{equation}
\label{eqn:K1a}
\frac{d}{d\rho}\left( \frac{\p \mathcal{L}}{\p \rho} \right) = \frac{d}{d\rho}\left(\frac{1}{2 g_{\rho\rho}} [g^2_{\rho \rho}(E^2 -V)]'\right) =  -\frac{1}{2}g_{\rho\rho}V''\,.
\end{equation}
Combining Eqs.~\eqref{eqn:K} and~\eqref{eqn:K1a}, the Lyapunov exponent is given by
\begin{equation}
\label{eqn:lya1}
\lambda = \sqrt{K_1 K_2} = \sqrt{-\frac{V''}{2\dot{t}^2}}\,.
\end{equation}
We recall that all quantities are evaluated at the unperturbed (circular) solution. With that in mind, we can arrive at an expression for $\lambda$ computed for $\rho=\rho_{\rm out}$, the outermost and unstable circular null geodesic. The conditions $V=E^2$ and $V'=0$ yield
\begin{equation}
\label{eqn:U_derivative}
\left( \frac{L}{E} \right)^2 = \rho^2 U^4 \text{\ \ and \ }U' = - \frac{U}{2\rho}\,.
\end{equation}
By substituting these expressions in Eq.~\eqref{eqn:lya1} and using the definition of $E$, we find:
\begin{equation}
\label{eqn:lya2}
\lambda =  \sqrt{\frac{-3U + 4\rho^2 U''}{2\rho^2 U^5}} \Bigg{|}_{\rho=\rho_{\rm out}}\,.
\end{equation}
It is possible to solve Eq.~\eqref{eqn:perturbation_matrix} by applying the standard procedure to solve coupled differential equations with constant coefficients. The solution is:
\begin{align}
\label{eqn:perturbed_prho}
\delta p_{\rho}(t) &= \delta p_0 \cosh({\sqrt{K_1 K_2}t}) + \sqrt{\frac{K_1}{K_2}} \delta\rho_0 \sinh(\sqrt{K_1 K_2}t), \\
\label{eqn:perturbed_rho}
\delta \rho(t) &= \delta \rho_0 \cosh({\sqrt{K_1 K_2}t}) + \sqrt{\frac{K_2}{K_1}} \delta p_0 \sinh(\sqrt{K_1 K_2}t)\,,
\end{align}
where $\delta p_0=\delta p_{\rho}(0)$ and $\delta\rho_0 =\delta\rho(0)$. We can compute $\lambda$ numerically from the previous expressions. To simplify the procedure, we set $\delta p_0 = 0$ and find that
\begin{equation}
 \label{eqn:lyapunov}
 \lambda = \frac{1}{t} \cosh^{-1}\left( \frac{\delta \rho(t)}{\delta \rho_0} \right) \
         = \frac{1}{t} \cosh^{-1}\left( \frac{\rho_{\rm num}(t) - \rho_{\rm out}}{\delta \rho_0} \right)\,,
\end{equation}
where $\rho_{\rm num}(t)$ is the solution of the radial equation found with a standard numerical integration technique.

\bibliography{references}

\end{document}